\documentclass[
 reprint,
 amsmath,amssymb,
jcp,
]{revtex4-1}
\usepackage{graphicx}
\usepackage{dcolumn}
\usepackage{bm}
\usepackage[usenames,dvipsnames,table,xcdraw]{xcolor} 
\usepackage[
	open,
	openlevel=2,
	atend
]{bookmark}
\usepackage[]{hyperref}            
\usepackage{tikz}
\usetikzlibrary{arrows}
\usepackage{flowchart}
\usepackage[version=4]{mhchem}
\usepackage{longtable}

\newcommand{\tf}[1]{\text{#1}}

\newcommand{\ti}[1]{\textit{#1}}
\newcommand{\etal}{\textit{et al.}~}
\newcommand{\etalcite}{\textit{et al.}}

\begin{document}
\newcommand{\addriChEM}{Collaborative Innovation Center of Chemistry for Energy Materials}
\newcommand{\addrSHMCIM}{Shanghai Key Laboratory of Molecular Catalysis and Innovative Materials}
\newcommand{\addrMOECPS}{MOE Key Laboratory of Computational Physical Sciences, Shanghai Key Laboratory of Bioactive Small Molecules}
\newcommand{\addrSKLBSM}{Shanghai Key Laboratory of Bioactive Small Molecules}
\newcommand{\addrFudan}{Department of Chemistry, Fudan University, Shanghai 200433, People's Republic of China}
\newcommand{\addrHFNL}{Hefei National Laboratory, Hefei 230088, China}
\newcommand{\addrZJLab}{Research Center for Intelligent Supercomputing, Zhejiang Lab, Hangzhou 311100, P. R. China}
\author{Sheng Bi}
\affiliation{ \addrFudan, \addrZJLab}
\author{Shirong Wang}
\affiliation{ \addrFudan}
	
\author{Igor Ying Zhang}
\email{igor\_zhangying@fudan.edu.cn}
\affiliation{ \addriChEM, \addrSHMCIM, \addrMOECPS, \addrSKLBSM, \addrFudan, \addrHFNL}
\author{Xin Xu}
\email{xxchem@fudan.edu.cn}
\affiliation{ \addriChEM, \addrSHMCIM, \addrMOECPS, \addrSKLBSM, \addrFudan, \addrHFNL}
\title{Improving XYG3-type Doubly Hybrid Approximation using Self-Interaction Corrected SCAN Density and Orbitals via the PZ-SIC Framework: the xDH@SCAN(SIC) Approach}
	
\begin{abstract}
XYG3-type doubly hybrid approximations (xDH) have gained a widespread recognition for their accuracy in describing a diverse range of chemical and physical interactions.
However, a recent study (\textit{J. Phys. Chem.} \textbf{2021}, \textit{12, 800-807}) has highlighted the limitation of xDH methods in calculating the dissociation of the NaCl molecule. 
This issue has been related to the density and orbitals used for evaluating the energy in xDH methods, 
which are obtained from lower-rung hybrid density functional approximations (DFAs) and display substantial density errors in the dissociation limit. 
In this work, we systematically investigate the influence of density on several challenging datasets and find that the xDH methods are less sensitive to the density errors compared to semi-local and hybrid DFAs. Furthermore, we demonstrate that the self-interaction corrected SCAN density offers superior accuracy 
compared to the self-consistent SCAN density and Hartree-Fock (HF) density, 
as evidenced by the charge analysis on the dissociation of heterodimers, such as NaCl and LiF. 
Building on these insights, we propose a 5-parameter xDH method using the SCAN density and orbitals corrected by the PZ-SIC scheme. 
This new xDH@SCAN(SIC) method provides a balanced and accurate description across a wide range of 
challenging systems.
\end{abstract}
\maketitle
\section{Introduction}

In Kohn-Sham (KS) \cite{cite-GEA-2} density functional theory (DFT) \cite{cite-GEA-1}, the central task is to find more and more accurate density functional approximations (DFAs) to the exchange-correlation (XC) functional. Doubly hybrid approximations (DHAs), which have become popular in quantum chemistry over the past decade \cite{first-ddhybrid, ddh-2, ddh-3,ddh-4, xDH-nonbond-pi,xDH-nonbond,ddh-5,xDH-fast}, 
incorporate virtual orbitals in the form of second-order perturbation (PT2) for the correlation energy in addition to occupied orbitals in the form of Hartree-Fock (HF) for the exchange energy. 
DHAs provide a consistent improvement over popular semi-local and hybrid DFAs, which are based solely on occupied orbitals. 
XYG3 is a typical DHA proposed by Zhang, Xu, and Gorddard in 2009 \cite{XYG3}, which defines a family of DHAs (xDH). 
Unlike other kinds of DHAs, xDHs insist on the use of self-consistent field (SCF) orbitals from a conventional lower-rung DFA to achieve a high-quality density, while constructing higher-rung functional for the purpose of achieving more accurate energy predictions.  
For instance, the xDHs of XYG3 and XYGJ-OS were established based on the B3LYP density and orbitals, while xDH-PBE0 used the reference density and orbitals from PBE0 calculations. 
For the main-group chemistry, several of the top-performing DFAs follow the xDH approach, including $\omega$B97M(2) proposed by the Head-Gordon group \cite{B97M}, xrevDSD-PBEP86-D4 by the Martin group \cite{xrevDSD}, and our own XYG7 \cite{xDH-GMTKN55}. 

A desired DFA should provide both accurate energy and density \cite{cite-GEA-2,DFT-straying,DFT-latest-2022}. 
However, enhancing energy accuracy does not necessarily ensure improvements in density,
particularly for DFAs that rely heavily on fitting to a broad range of energy data with no concern of density data \cite{DFT-straying}.
In the context of density-corrected DFT (DC-DFT), Burke and co-workers \cite{inconsistent,dc-improved,dc-theory} categorized the errors as either density-driven or functional-driven. They showed that there are many density-sensitive cases where the impact of the density-driven errors cannot be ignored \cite{density-error}. 
The interplay between these two types of errors poses a key challenge in functional development, 
especially for density-sensitive systems \cite{dc-improved}. 
Interestingly, the work of Burke \etalcite \cite{inconsistent}, Scuseria \etalcite \cite{dc-hf-obt-reaction}, and Bartlett \etalcite \cite{cite-reaction-ccsdt} has consistently shown that substituting self-consistent densities from DFAs with HF densities can be sufficient. 
This observation aligns with the development philosophy of xDH methods \cite{XYG3,lrc-XYG3,xDH-GMTKN55,cite-ddhybrid-MSE}, 
suggesting that higher accuracy in energy can be achieved without invoking a full self-consistency in density determination. 

The dissociation of NaCl serves as a prototypical example of density-sensitive scenarios that highlights a significant error in densities, particularly in the stretched-bond region \cite{cite-SIC-intro, cite-DC}. 
These errors can be attributed to the substantial self-interaction errors (SIE) \cite{cite-SIC-intro,cite-SIC-4-manybody,cite-DFT-SIE}, 
or delocalization errors \cite{cite-delocalized-error}, present in popular semi-local or hybrid DFAs \cite{cite-SIC-SCAN, Bi2023Self-interaction,cite-sie-ddhybrid}, such as meta-GGA SCAN \cite{cite-scan-1}, hybrid B3LYP \cite{b3lyp}, and PBE0 \cite{cite:pbe0}. 
These functionals exhibit a preference for densities with partially occupied (or delocalized) electrons on each nucleus, deviating significantly from the actual neutral state in the dissociation limit \cite{dc-improved}. 
Despite their success in numerous systems of chemical and physical interest \cite{cite-ddhybrid-MSE, scsRPA,xdh-RPA,xDH-nonbond,xDH-nonbond-pi}, xDH methods, 
such as XYG3 and xDH-PBE0, still carry noticeable density-driven errors inherited from the reference density of B3LYP and PBE0, respectively \cite{xDH-incorrect}. 
Suhwan \etal found that XYG3 predicts an incorrect dissociation limit for NaCl \cite{DC-HF-ddhybrid}. 
Moreover, they demonstrated that using HF densities effectively mitigates the density-driven errors in xDHs. 
The resulting one-parameter XYG3-type doubly hybrid BL1p dissociates NaCl correctly (see Fig. 6 in Ref. \cite{DC-HF-ddhybrid}).

Recently, we have demonstrated that the xDH methods can produce accurate density through external potential perturbation \cite{xDH-pot-pert}.
Consequently, when utilizing the xDH relaxed densities, it is anticipated that the density-driven errors in xDH can be mitigated, 
albeit at the cost of increased computational expenses.
It is important to note that HF densities, which lack the correlation effect, are generally considered inferior to the self-consistent densities of popular semi-local and hybrid DFAs \cite{cite-SIC-intro,cite-SIC-gap,cite-PBEh,cite-SIC-2}. The positive role of using HF densities in alleviating the density-driven errors in DFAs can be attributed, at least in part, 
to the tendency of the HF method to over-localize electron density. 
This over-localization tendency contrasts with the delocalization tendency of semi-local and hybrid DFAs, 
which enables HF densities to capture the correct dissociation limit qualitatively.

To what extent can the performance of xDH methods be further improved by using more accurate densities? 
The answer to this question depends on various factors. 
While the HF density is commonly used in DC-DFT, an alternative, which is worthy of exploring, is 
the self-interaction corrected (SIC) density of semi-local DFAs.
The SIC method, initially proposed by Perdew and Zunger (PZ) in the early 1980s, aims to address the SIE by incorporating one-electron correction terms into the semi-local DFAs \cite{cite-SIC-4-manybody}. 
The SIC density is superior to the HF density in the sense that the correlation effect has been properly counted from the semi-local correlation functional. 
Moreover, previous studies have demonstrated that the SIC scheme not only precisely removes the one-electron SIE, but also significantly relieves the many-electron SIE in the semi-local DFAs \cite{cite-SIC-intro}. 

In this paper, we establish a new xDH model with 5 empirical parameters (xDH@DFA, Eq.~(\ref{eq:xDH@DFA})) and propose two new xDH methods, namely xDH@SCAN and xDH@SCAN(SIC). 
They share the same functional formula (Eq.~\ref{eq:xDH@SCAN}), but use different reference densities and orbitals.
To be specific, xDH@SCAN utilizes the self-consistent density and orbitals from the state-of-the-art meta-GGA SCAN,
while xDH@SCAN(SIC) employs the self-interaction corrected SCAN (SCAN(SIC)) density using the PZ-SIC algorithm. 
We assess the performance of these xDHs on the SIE4x4 test set \cite{cite-GMTKN55-detail}, which contains four positively charged homodimers of H$_2^{+}$,
He$_2^{+}$, $(\ce{NH3})_2^+$, and $(\ce{H2O})_2^+$.
These cases exhibit heavy SIEs in semi-local DFAs, although they are known to be not density-sensitive.
As a result, substituting the self-consistent density with the HF density essentially does not improve the performance of semi-local DFAs.
Consequently, we investigate 76 barrier heights (the BH76 database \cite{cite-bh76-1,cite-bh76-2}), where semi-local DFAs significantly underestimate the values.
These transition states are considered density-sensitive and the use of HF density improves the performance of semi-local DFAs  \cite{dc-hf-obt-reaction}.
Remarkably, our results demonstrate that both the SIE4x4 and BH76 databases could be accurately described within the xDH framework, 
regardless of whether the PZ-SIC correction is applied to the reference densities or not. 
This observation suggests that the density-driven error is much less pronounced in xDH methods compared to lower-rung DFAs.

While it would be interesting to make a direct comparison between DFA densities and some numerically exact references, there is a lack of a universally applicable mathematical measure for directly assessing density errors in DFT calculations \cite{quantify_density_error}. Here, we present the dissociation curves of LiF and NaCl as our prototypical systems, where the density-driven error becomes noticeable for xDH methods.
The previous study by Suhwan \etal (Fig. 6 in Ref. \cite{DC-HF-ddhybrid}) reported the poor performance of XYG3 for NaCl. 
Our finding reveals that another xDH method based on PBE0 density and orbitals, i.e. xDH-PBE0\cite{xdhpbe0}, exhibits a similar inadequate performance. 
In contrast, utilizing the PZ-SIC corrected SCAN density in xDH@SCAN(SIC) effectively eliminates these density-driven errors and yields correct dissociation curves for both molecules. 
Furthermore, it has been found that even though the HF density improves the method performance in density-sensitive scenarios, it often deteriorates in density-insensitive cases. 
Our results demonstrate that by utilizing better densities obtained from the PZ-SIC corrected SCAN method, xDH@SCAN(SIC) achieves comparable accuracy to xDH@PBE0 and xDH@SCAN in the systems with fewer density-driven error as evidenced by the benchmark of the atomization energies in the G2-1 and W4-11 databases, as well as the reactions involving alkaline compounds in the ALK8 database.

\section{Method} \label{sec:method}
In the current study, we examine another xDH functional, namely xDH-PBE0, which utilizes the density and orbitals from the 
nonempirical hybrid PBE0 functional\cite{xdhpbe0}.
The xDH-PBE0 functional can be expressed in a general form that incorporates 5 parameters, denoted as xDH@DFA \cite{cite-sie-ddhybrid}:
\begin{equation} \label{eq:xDH@DFA}
  \begin{aligned}
     E_{xc}^{\tf{xDH@DFA}} = &c_1 E_x^{\tf{HF}} + c_2 E_x^{\tf{DFA}} + c_3 E_c^{\tf{DFA}} \\
    & + c_4 E^{\tf{osPT2}}+ c_5 E^{\tf{ssPT2}}  \quad .
  \end{aligned}
\end{equation}
Here, $E_x^{\tf{HF}}$ denotes the HF-like exchange.
$E^{\tf{osPT2}}$ refers to the opposite-spin component of the PT2 correlation  (osPT2), while $E^{\tf{ssPT2}}$ signifies the same-spin component of the PT2 correlation (ssPT2) \cite{PT2}.  
Within the xDH@DFA model, $E_x^{\tf{DFA}}$ and $E_c^{\tf{DFA}}$ correspond to the exchange and correlation contributions from a specific lower-rung DFA that is employed to generate 
the reference density and orbitals for the final xDH energy calculations. 
Given that the PBE0 density and orbitals are employed, xDH-PBE0 (also named xDH@PBE0 in this work) incorporates the PBE exchange and correlation terms in its formula ($E_x^{\tf{PBE}}$ and $E_c^{\tf{PBE}}$) \cite{xdhpbe0}.
Meanwhile, by applying two extra constraints,
\begin{equation}
  \begin{aligned}
    & c_2 = 1.0 - c_1, \quad  c_5 = 0.000,\quad
  \end{aligned}
\end{equation}
xDH@PBE0 contains 3 empirical parameters that were optimized against the errors in the calculated heats-of-formation for the G3/99 dataset,
\begin{equation} \label{eq:xDH@PBE0-coeff}
  \begin{aligned}
    & c_1 = 0.8335, \quad  c_3 = 0.5292, \quad c_4 = 0.5428. \quad \\
  \end{aligned}
\end{equation}

To further investigate the influence of density errors on the final xDH performance, we started with the aforementioned general form of xDH@DFA and developed a new xDH based on the reference density and orbitals from the cutting-edge meta-GGA functional SCAN \cite{cite-scan-1}, which we refer to as xDH@SCAN: 
\begin{equation} \label{eq:xDH@SCAN}
  \begin{aligned}
     E_{xc}^{\tf{xDH@SCAN}} = &c_1 E_x^{\tf{HF}} + c_2 E_x^{\tf{SCAN}} + c_3 E_c^{\tf{SCAN}} \\
    & + c_4 E^{\tf{osPT2}}+ c_5 E^{\tf{ssPT2}}  \quad.
  \end{aligned}
\end{equation}
The non-empirical SCAN contains 17 parameters, all of which were determined by known exact constraints. We then optimized the 5 parameters of xDH@SCAN by utilizing three datasets of SIE4x4, BH76, and G2-1,
\begin{equation}\label{eq:xDH@SCAN-coeff}
  \begin{aligned}
    & c_1 = 0.822,\quad  c_2 = 0.204,\quad \\
    & c_3 = 0.543,\quad  c_4 = 0.385,\quad c_5 = 0.160 \quad. \\
  \end{aligned}
\end{equation}
The datasets and the performance of xDH@SCAN will be discussed below.

Recently, an efficient PZ-SIC implementation was proposed and implemented in FHI-aims \cite{Bi2023Self-interaction}. 
A self-consistent localization strategy was established to corporate with the standard constraints of the PZ-SIC method. 
The comprehensive benchmark has demonstrated that this strategy can help to effectively find a feasible set of SIC orbitals and eventually achieve the minimal PZ-SIC energy
\begin{equation} \label{eq:SIC-operator}
  \begin{aligned}
  E^{\tf{SIC-SCAN}} = E^{\tf{SCAN}}
  - \sum_i^{N_{\text{e}}} f_i \left( E_{\text{xc}}^{\text{SCAN}}[n_{i}]+E_{\text{H}}[n_{i}] \right) \quad.
  \end{aligned}
\end{equation}
Here $ \{ n_i(\textbf{r}) \}$ are the densities of the $N_e$ single-electron orbitals,
called SIC orbitals. Accordingly, $\{ f_i \}$ are the occupation numbers for the SIC orbitals \cite{cite-SIC-1}. 
$E_{\text{xc}}^{\text{SCAN}}[n_{i}]$ is the self-exchange-correlation energy evaluated by the orbital density of $n_i(\textbf{r})$ for the SCAN exchange-correlation functional, 
while $E_{\text{H}}[n_{i}]$ is the self-Hartree energy for~$n_i(\textbf{r})$. 
For more details about the PZ-SIC implementation in FHI-aims, interesting readers are referred to Ref.~\cite{Bi2023Self-interaction}.
In this work, we aime to explore how density correction via the PZ-SIC scheme can enhance the performance of xDH. 
With this in mind,
we proposed the xDH@SCAN(SIC) functional, which follows the same formula as xDH@SCAN (Eq.~\ref{eq:xDH@SCAN}). 
The only difference is using the self-interaction corrected SCAN density and orbitals via PZ-SIC algorithm.
The 5 parameters in xDH@SCAN(SIC) were then re-optimized against the same datasets as for xDH@SCAN
\begin{equation}\label{eq:xDH@SCANSIC-coeff}
  \begin{aligned}
    & c_1 = 0.717,\quad    c_2 = 0.306, \quad  \\
    & c_3 = 0.560,\quad    c_4 = 0.363, c_5 = 0.092. \quad \\
  \end{aligned}
\end{equation}

\section{Results and discussion} \label{sec:DIS}

The SIE is a challenging issue for widely used semi-local and hybrid DFAs \cite{cite-SIC-4-manybody,cite-SIC-2}. 
A classic example of the SIE is the dissociation limit of positively charged homodimers (A$\cdots$A)$^+$, 
where the monomer A can comprise atoms, such as H and He, or molecules, such as $\ce{NH3}$, and $\ce{H2O}$. 
By definition, the state with symmetric charge A$^{+0.5}\cdots$A$^{+0.5}$ should be degenerate in energy with symmetry broken states A$^{+q}\cdots$A$^{1-q}$ with $0\le q\le 1$. 
The SIE is also referred as the delocalization error, stemming from its strong inclination towards the symmetric charge state with $q=0.5$ \cite{cite-SIC-2}.

\begin{table*}[]
  \caption{ 
    Mean absolute errors (MAEs, in kcal/mol) of different methods against the datasets of SIE4x4, BH76, HD2x5, G2-1, W4-11, and ALK8.
    The corresponding maximum absolute errors (Max) are provided in parentheses.
  }
  \label{table-mae}
  \begin{tabular}{|c|rr|rr|rr|rr|rr|rr|}
    \hline
    \textbf{DFAs} & \multicolumn{2}{c|}{\textbf{SIE4x4}} & \multicolumn{2}{c|}{\textbf{BH76}} & \multicolumn{2}{c|}{\textbf{G2-1}} & \multicolumn{2}{c|}{\textbf{W4-11}} & \multicolumn{2}{c|}{\textbf{ALK8}} & \multicolumn{2}{c|}{\textbf{HD2x5}} \\ \hline
    SCAN          & 18.12 & (40.38)  & 7.74  & (18.10)  & 3.35   & (10.87)   & 3.53  & (26.03)  & 3.13  & (8.36) & 4.09   & (10.25)   \\
    SCAN@HF       & 14.18 & (34.96)  & 3.48  & (15.22)  & 5.16   & (19.69)   & 6.35  & (31.02)  & 2.06  & (5.16) & 8.62   & (27.50)   \\
    SCAN(SIC)     & 17.90 & (39.63)  & 5.49  & (11.04)  & 3.45   & (10.44)   & 2.92  & (13.17)  & 3.16  & (7.08) & 2.64   & (6.58)   \\
    xDH@PBE0      & 1.10  & (2.26)  & 1.30  & (4.25)  & 2.91   & (10.00)   & 3.60  & (16.29)  & 2.04  & (6.80) & 19.25  & (83.33)   \\
    xDH@SCAN      & 0.53  & (1.23)  & 1.14  & (5.43)  & 3.47   & (18.18)   & 4.33  & (29.55)  & 5.03  & (19.46) & 18.55  & (50.16)   \\
    xDH@SCAN(SIC) & 1.95  & (4.15)  & 1.98  & (9.52)  & 2.89   & (15.00)   & 3.16  & (25.64)  & 2.98  & (10.54) & 2.24   & (5.61)    \\ \hline
    \end{tabular}
\end{table*}

\begin{figure}[!htbp] 
  \vspace{0.0in}
  \begin{center}
    \includegraphics[width=3.2in]{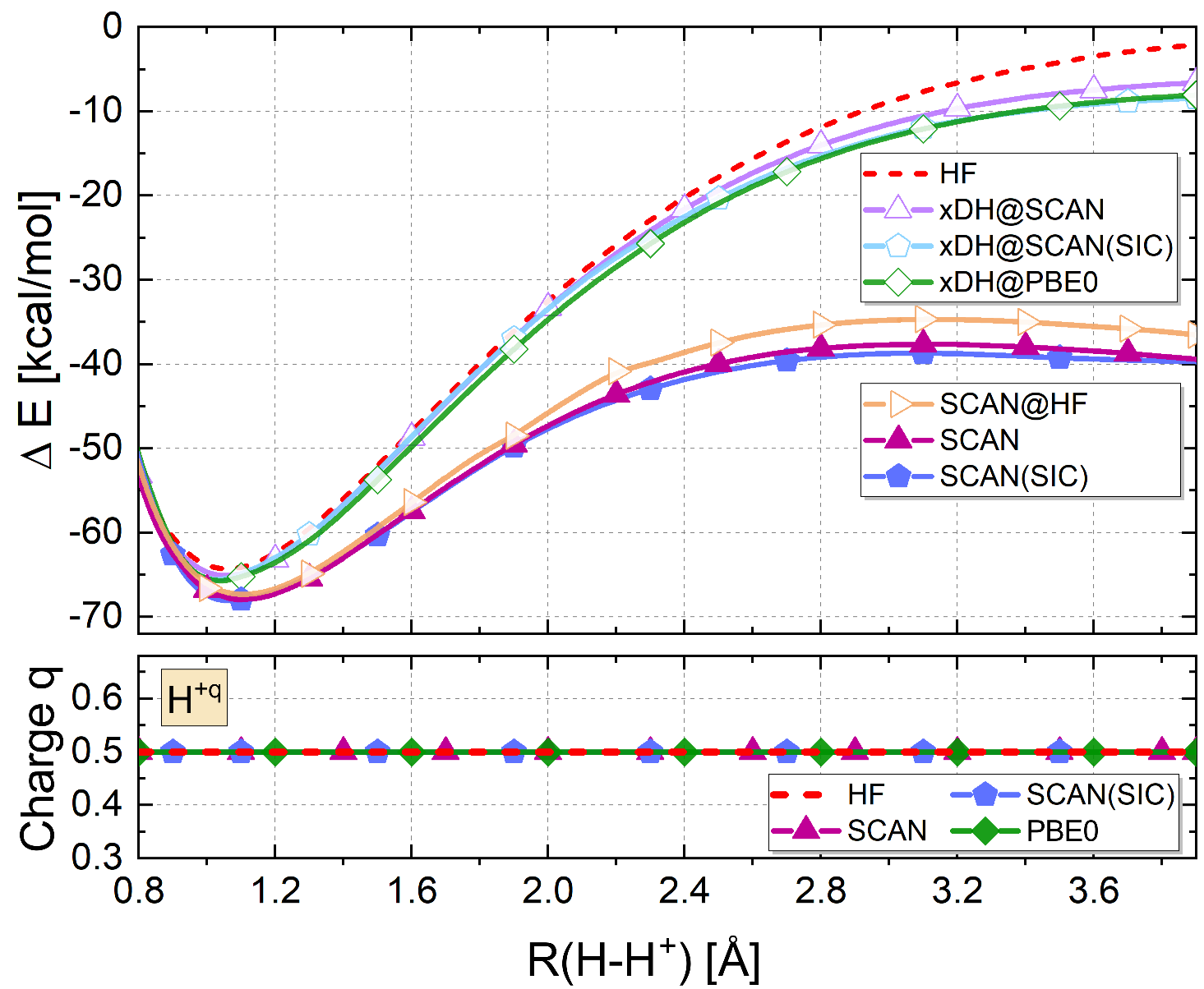}
    \end{center}
  \vspace{-0.2in}
  \caption{ 
      (top) H$_2^+$ dissociation curves calculated by various methods. The total energies of isolated atoms/ions are used as the zero energy level.
      (bottom) Charge analyses along the dissociation, where the Hirshfeld charges on the H nucleus are provided. 
      All results were obtained with the \ti{tight} basis sets using the \ti{FHI-aims} package \cite{cite-aims-1}.
            }
  \label{Fig:SIE-dissociation}
  \end{figure}

\begin{figure}[!htbp] 
  \vspace{0.0in}
  \begin{center}
  \includegraphics[width=3.4in]{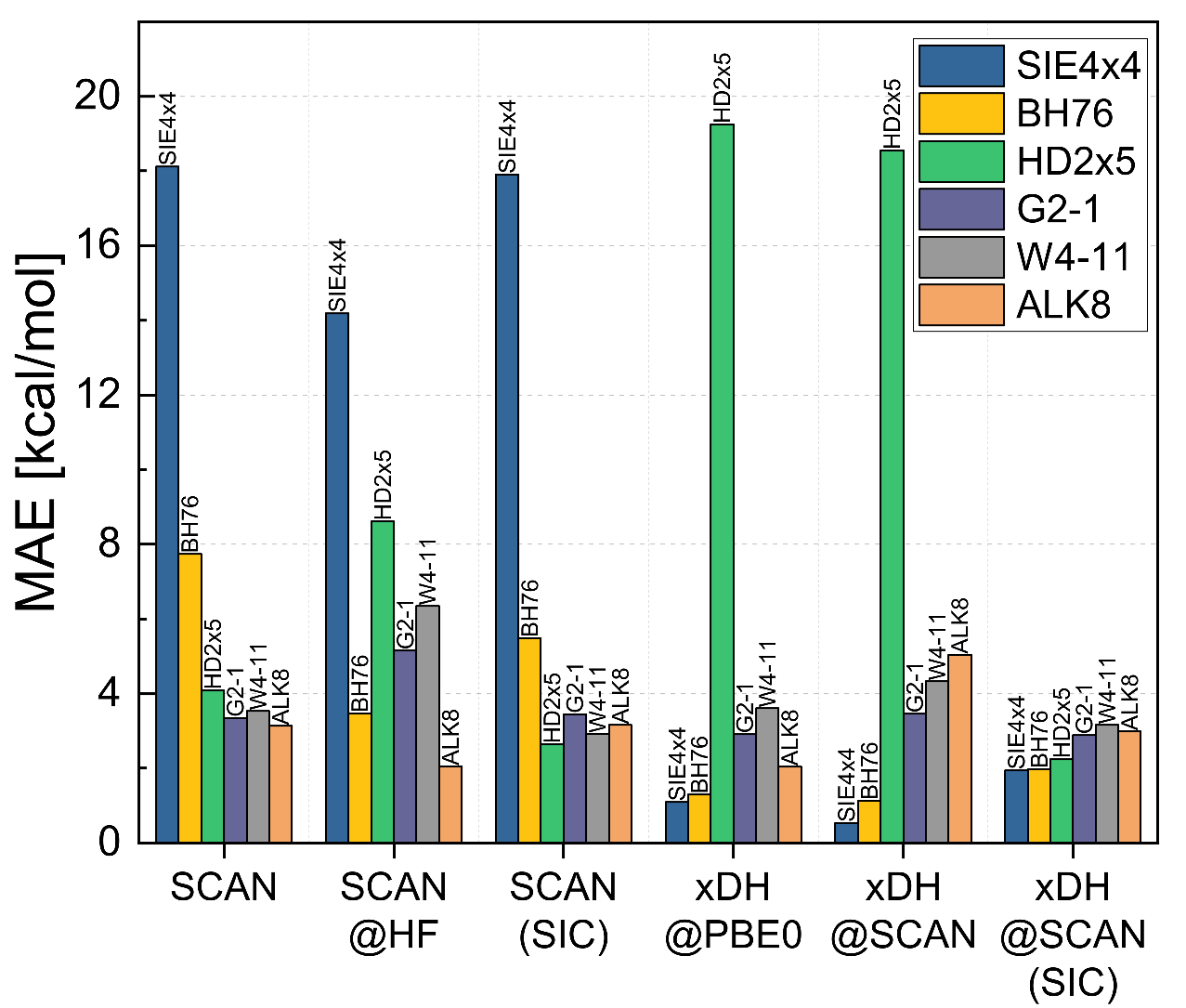}
    \end{center}
  \vspace{-0.2in}
  \caption{ 
    Mean absolute errors (MAEs) of different DFAs against various datasets, including SIE4x4, BH76, HD2x5, G2-1, W4-11, and ALK8. 
    The reference data for SIE4x4, BH76, G2-1, W4-11 and ALK8 are taken from Ref. \cite{cite-GMTKN55-detail}, 
    while the accurate CCSD(T) results are used for the HD2x5 dataset. 
    All calculations are performed with the \ti{tight} basis set using the \ti{FHI-aims} package. 
      }
  \label{Fig:mae}
  \end{figure}

Figure~\ref{Fig:SIE-dissociation} presents the H$_2^+$ dissociation curves of various DFAs. 
The HF method, which is exact for any one-electron system, serves as the reference for other dissociation curves. 
In our plot, the zero energy level is set to the total energy of the symmetry broken state ($q=1$) in the dissociation limit,
while the charge symmetry ($q=0.5$) is constrained throughout the entire dissociation curve, as shown at the bottom of 
Figure~\ref{Fig:SIE-dissociation}.
Due to the degeneracy of different charge states in the dissociation limit, 
the correct dissociation curve, as provided by the HF method, tends to zero in the dissociation limit.
The SIE in semi-local DFAs, as represented by SCAN in this work, quickly deteriorates the functional performance when stretching the H-H bond away from the equilibrium, leading to errors by more than 30 kcal/mol in the dissociation limit.

As shown in Figure~\ref{Fig:SIE-dissociation}, using different densities does not notably change 
the functional performance. The SCAN dissociation curves using the HF density and the PZ-SIC corrected SCAN density are denoted as SCAN@HF and  SCAN(SIC), respectively. 
In line with previous literature \cite{inconsistent}, our results confirm that positively charged homodimers are not density-sensitive, where the SIE of DFAs are primarily functional-driven.
In consequence, a much better performance can is obtained by using the XYG3-type doubly hybrid methods of xDH@PBE0, xDH@SCAN, and xDH@SCAN(SIC), where the SIE is small because of the high portion of exact exchange employed in the xDH functionals. 

To further study the SIE in positively charged homodimers, we benchmark the performance of various methods against the SIE4x4 dataset, which contains 4 data points along each dissociation curve of H$_2^{+}$, He$_2^{+}$, $(\ce{NH3})_2^+$ and $(\ce{H2O})_2^+$. 
The 4 data points are sampled at $1.0$ times, $1.25$ times, $1.5$ times, and $1.75$ times the respective equilibrium distance. 
The mean absolute errors (MAEs) of SCAN, SCAN@HF, SCAN(SIC), 
and the corresponding xDH methods for SIE4x4 are given in Table~\ref{table-mae}, which is also presented in Figure~\ref{Fig:mae} for convenience.
We see again that changing the reference density is unable to alleviate the large SIEs in semi-local DFAs. 
The MAEs of SCAN, SCAN@HF, and SCAN(SIC) in SIE4x4 are all above 10 kcal/mol. 
For comparison, the xDH methods perform much better. 
The MAEs are 1.10 kcal/mol, 0.53 kcal/mol, and 1.95 kcal/mol for xDH@PBE0, xDH@SCAN, and xDH@SCAN(SIC), respectively.

We now examine the functional performance on reaction barrier heights. 
It is well-known that semi-local DFAs tend to underestimate the reaction barrier heights. This error has often been attributed to the SIEs in the semi-local DFAs \cite{cite-reaction-DFA}, 
based on the observation that this underestimation can be effectively alleviated by introducing a portion of exact exchange in forming a hybrid DFA.
Meanwhile, Scuseria \etalcite \cite{dc-hf-obt-reaction} and Bartlett \etalcite \cite{cite-reaction-ccsdt} found that replacing the self-consistent densities with HF densities can also improve the DFA performance consistently, 
and Burke \etal attributed the error of semi-local and hybrid DFAs in reaction barriers to density-driven \cite{dc-improved}.

\begin{figure*}[!htbp] 
  \vspace{0.0in}
  \begin{center}
      \includegraphics[width=3.4in]{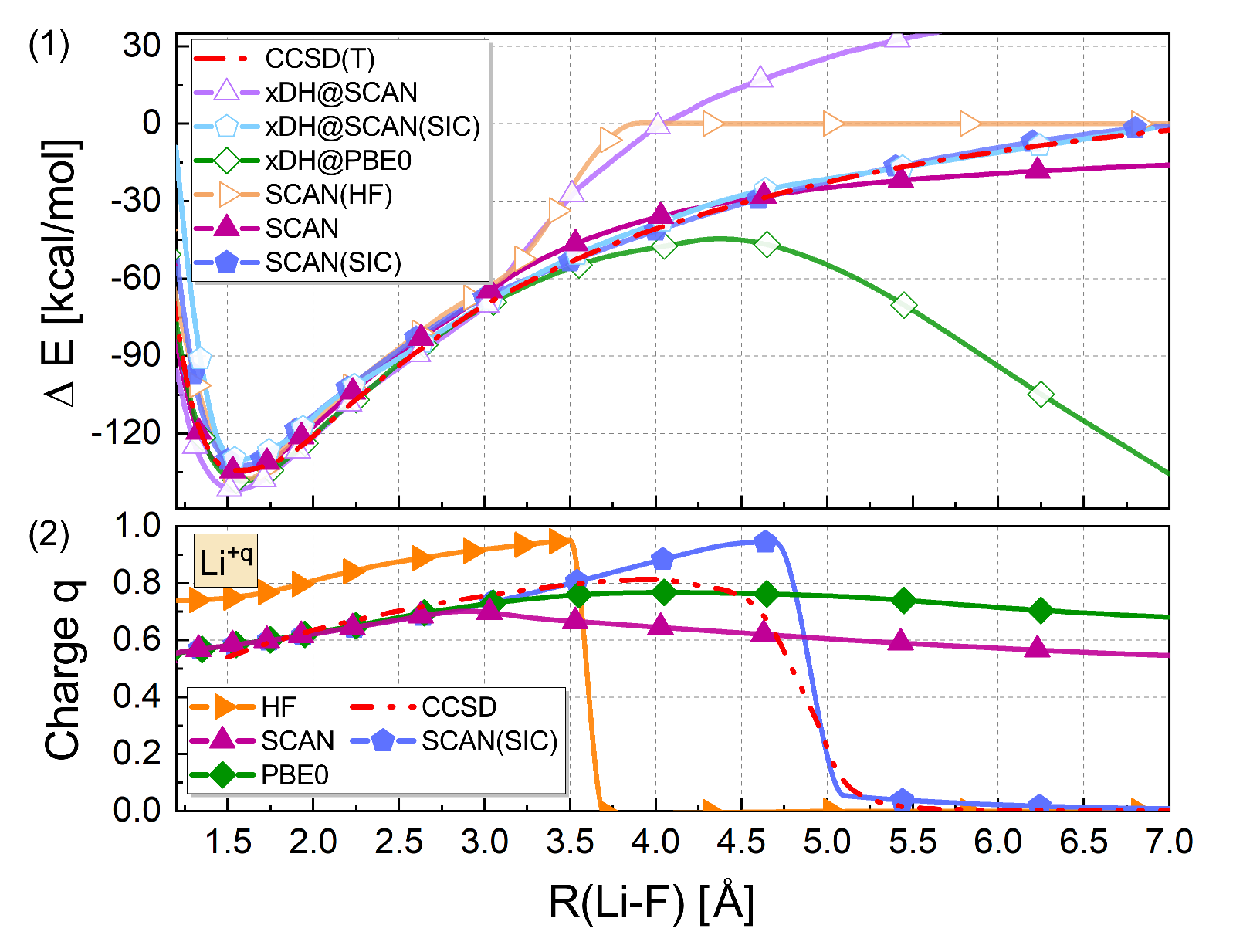}
  \includegraphics[width=3.4in]{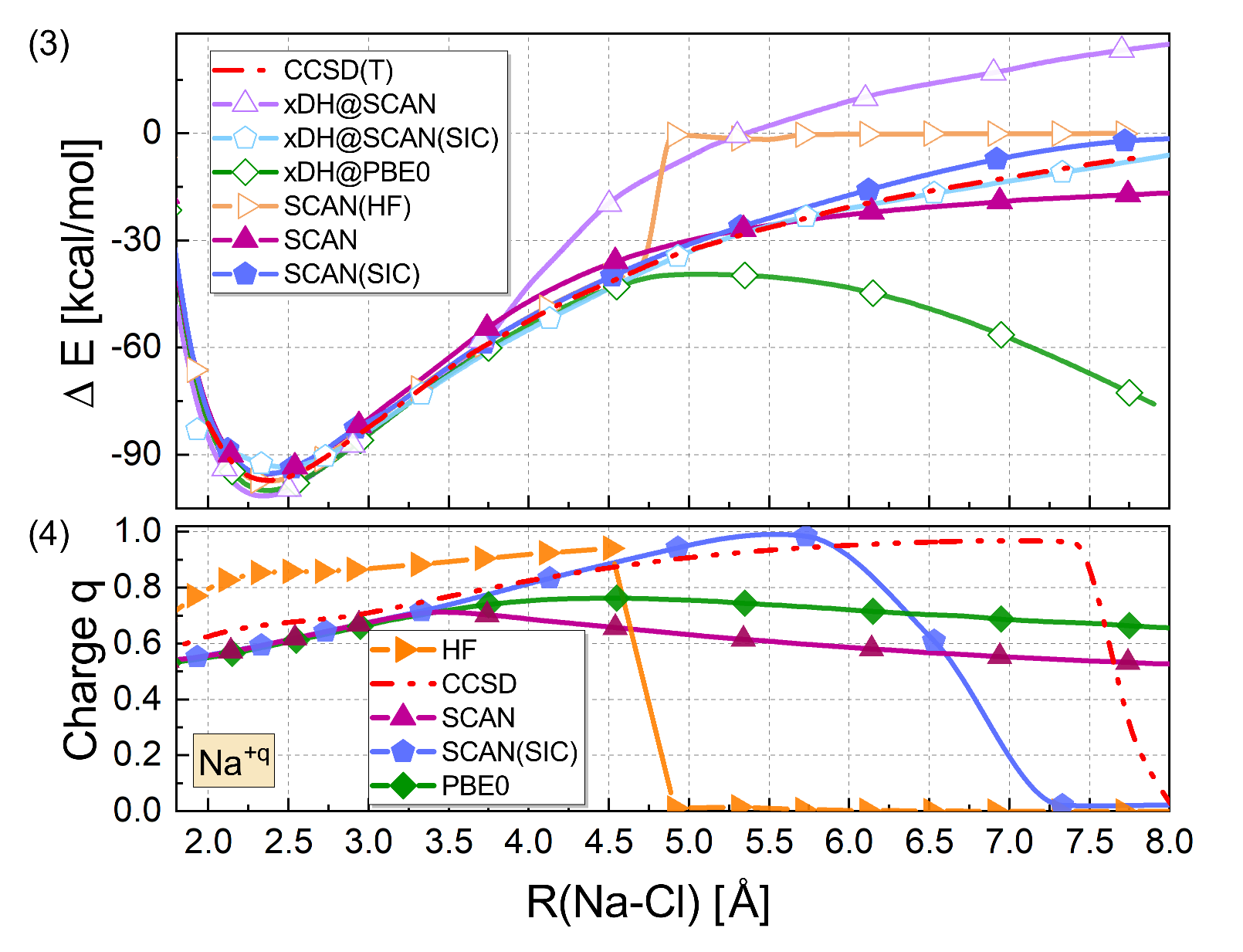}
  \end{center}
  \vspace{-0.2in}
  \caption{ 
      Dissociation curves of LiF (1) and NaCl (3) calculated by various DFAs. The zero energy level is set to the total energy of isolated atoms. 
      Charge analyses along the dissociation curves of LiF (2) and NaCl (4), where the Hirshfeld charges on Li and Na nucleus are provided, respectively.
      Most of the calculations are performed with \ti{tight} basis sets using the \ti{FHI-aims} package.
      The accurate CCSD(T) results are taken as the reference for both LiF \cite{cite-SIC-26-Yang-LSC} and NaCl \cite{DC-HF-ddhybrid} dissociation curves.
      The Hirshfeld charges computed at the CCSD level are used as the reference for the charge analyses, which are obtained using the \ti{Gaussian} package \cite{cite-Gaussian} at the basis set level of cc-pVTZ.
      }
    \label{Fig:Li-F-xDH}
\end{figure*}

The most widely used dataset for reaction barrier heights is BH76, which contains 76 forward and reverse barrier heights for a set of reactions including hydrogen-transfer, heavy-atom-transfer, nucleophilic-substitution, and unimolecules \cite{cite-bh76-1,cite-bh76-2}. 
The MAEs of various DFAs on BH76 are collected in Table~\ref{table-mae}. 
Our findings agree with the previous studies \cite{cite-reaction-DFA} that the semi-local SCAN functional consistently underestimates the reaction barrier heights in BH76, leading to a large MAE of 7.74 kcal/mol.
Without changing the functional, this error can be effectively reduced by using either HF density or PZ-SIC corrected SCAN density. 
The MAEs of SCAN@HF and SCAN(SIC) are 3.48 kcal/mol and 5.49 kcal/mol, respectively. 
All these findings identify the density-driven error of SCAN to be notable in BH76.
It is known that the xDH methods can provide very accurate description of reaction barrier heights \cite{xDH-GMTKN55}. 
The MAE of xDH@PBE0 is 1.30 kcal/mol, which is further reduced to only 1.14 kcal/mol when the SCAN exchange and correlation are employed in the 5-parameter xDH@DFA model and the self-consistent SCAN density and orbitals are utilized. 
Furthermore, our results show that the accuracy of xDH@DFA is largely maintained with different densities, while xDH@SCAN(SIC) results in an MAE of 1.98 kcal/mol.

Hence, when considering positively charged homodimers in SIE4x4 and reaction barrier heights of BH76 together, 
an entirely different influence of the reference densities is revealed for semi-local DFAs. 
Specifically, the SCAN error is identified as density-driven for BH76, because a notable improvement can be achieved if replacing the self-consistent SCAN density by HF or SCAN(SIC) densities.
However, using these densities cannot address the failure of SCAN on the SIE4x4 dataset indicating the SCAN error in this case is mainly functional-driven.  

In comparison, xDH methods perform significantly better on the SIE4x4 dataset, indicating a minor functional-driven error. 
Moreover, from the perspective of many-body perturbation theory, the second-order perturbative energy used in xDHs is associated with the first-order perturbative wave function. 
For most chemical and physical interest systems, the density error is small and cannot be negligible in real space. 
Standard perturbation theory is often used to improve density.
Consequently, xDH methods have partially corrected the density-driven error at the first-order perturbation level and thus exhibit less sensitivity to the choice of density against the BH76 dataset.

From this perspective, the failure of xDH methods in the NaCl dissociation, reported by Suhwan \etalcite~\cite{DC-HF-ddhybrid}, 
shall be traced back to the errors in the reference density of hybrid DFAs, which are too large to be corrected in a perturbative manner. 
Figure~\ref{Fig:Li-F-xDH} presents the dissociation curves of LiF and NaCl, both of which are the prototypical examples of heterodimers A$\cdots$B with the monomer A denoting the metal atoms, i.e., Li and Na, and B for halogen atoms of F and Cl. 
We also plotted the Hirshfeld charge $q$ on Li and Na nuclei along the dissociation. 
Unlike the positively charged homodimers discussed above, there is no mirror symmetry for LiF and NaCl, 
such that the charge distribution, i.e., $q$ in A$^{+q}\cdots$B$^{-q}$, changes with respect to the bond distance change. 
The error in predicting $q$ is the straightforward measurement of the corresponding density error. 
By taking the $q$ values at the CCSD level as the reference, 
we can see that the Hirshfeld charges at the equilibrium bond distance are around 0.6 for both LiF and NaCl. 
Both semi-local SCAN and hybrid PBE0 can provide accurate $q$ values and thus have small density errors in the equilibrium region. 

However, in the stretched-bond region, the SIE in SCAN and PBE0 becomes dominant and induces large 
errors in their self-consistent densities. 
As suggested by the CCSD results in Figure~\ref{Fig:Li-F-xDH}, charge $q$ evolves to zero when the stretched bond is fully dissociated. 
This shows no charge transfer between the metal and halogen atoms when they are well separated. 
Consequently, we set the zero energy for each dissociation curve in Figure~\ref{Fig:Li-F-xDH} to the total energy of neutral atoms calculated using the same method. 
It becomes evident that both SCAN and PBE0 favor a delocalized charge distribution, 
yielding non-zero $q$ values even in the dissociation limit. The incorrect density predicted by PBE0 and SCAN results in the inaccurate
dissociation tails of xDH@PBE0 and xDH@SCAN for NaCl and LiF.
It is worth noting that the XYG3 dissociation curve of NaCl, as reported by Suhwan \etalcite~\cite{DC-HF-ddhybrid}, 
exhibits a similar dissociation behavior as xDH@PBE0. This was attributed to the large density error in the B3LYP density.

For comparison, the HF method yields the correct charge distribution in the heavily stretched-bond region. 
However, the HF density in the equilibirum region is inaccurate, which significantly overestimates the $q$ values by approximately 0.2 for both heterodimers.
A more accurate density can be achieved using the self-interaction corrected SCAN density via the PZ-SIC algorithm. 
This SCAN(SIC) density proves to be accurate in both equilibrium and heavily stretched-bond regions,
demonstrating its superiority over the HF density in these scenarios.

The charge analysis depicted in Figure~\ref{Fig:Li-F-xDH} provides a clear illustration of the electron transfer for the metal-halogen heterodimers.
This process follows the so-called harpooning mechanism \cite{harpooning-mechanism, harpooning-AlO, Herschbach-harpooning}. 
This mechanism is characterized by the existence of a specific electron transfer distance at which neutral-ionic avoided crossing takes place. 
Around this particular distance, an abrupt electron transfer between Na (or Li) and Cl (or F) is observed. 
Our results suggest that semi-local SCAN and hybrid PBE0 completely fail in predicting this electron-transfer distance. 
Although the HF method yields an electron-transfer distance, 
it is too closer to the equilibrium region. 
The density error in this region is partly responsible for the abnormal dissociation behavior of the SCAN(HF) method, 
which uses the HF density as the reference. 
Again, the SCAN(SIC) density is more accurate and provides the electron transfer profile that is closer to that of the CCSD method.
Using the accurate SCAN(SIC) density and orbitals, xDH@SCAN(SIC) reproduces accurately the dissociation curves for both NaCl and LiF,
which overlaps with the CCSD(T) reference curves precisely.

\begin{figure}[!htbp] 
    \begin{center}
    \includegraphics[width=3.4in,angle=0]{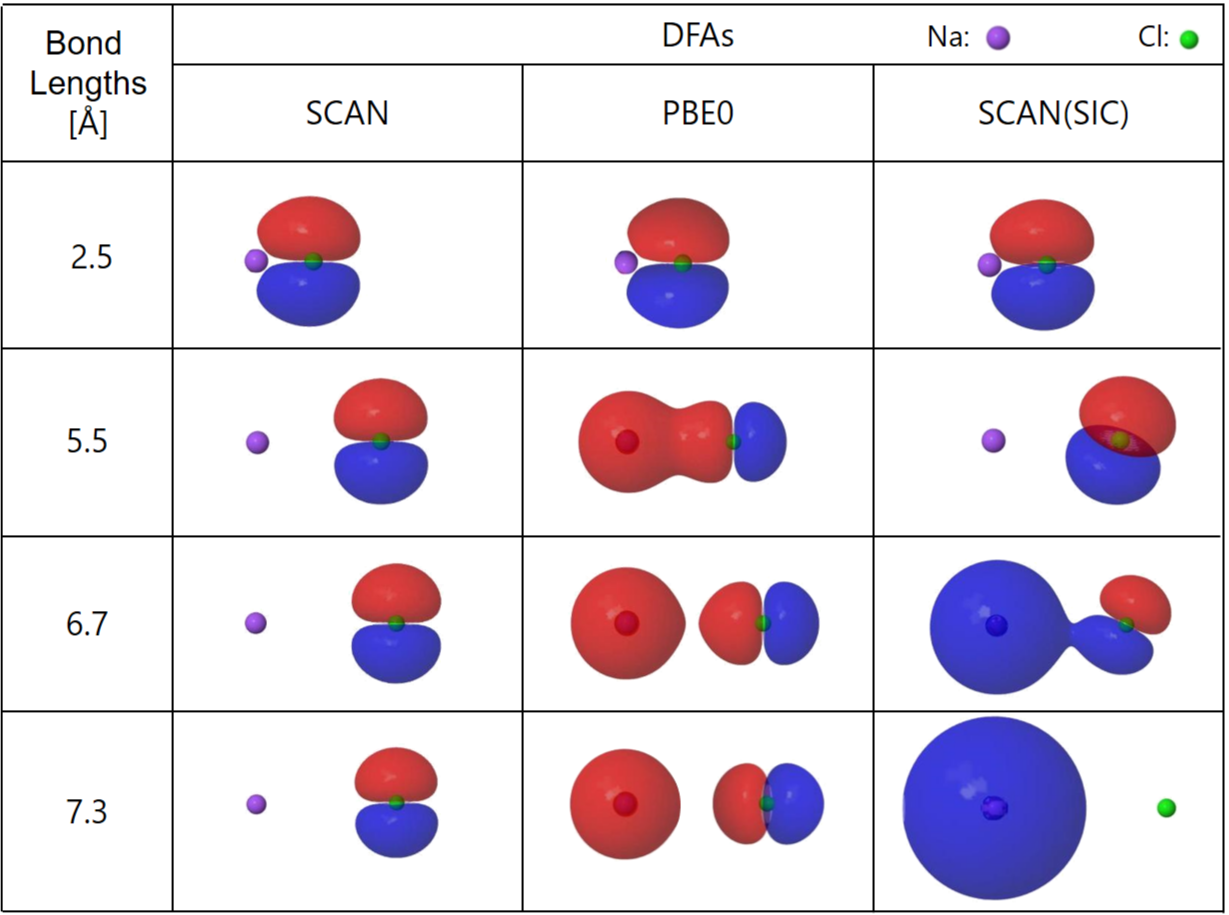}
      \end{center}
    \caption{ 
    Highest occupied molecular orbitals (HOMOs) calculated by SCAN, PBE0, and SCAN(SIC) methods for NaCl molecules with the bond lengths 
    at $2.5$ \AA, $5.5$ \AA, $6.7$ \AA, and $7.3$ \AA, respectively.
    All calculations were performed with the \ti{tight} basis sets provided by \ti{FHI-aims}. 
  }
  \label{Fig-NaCl-cube}
  \end{figure}

To gain a deeper insight of the electron transfer process in these heterodimers, we have depicted the highest occupied molecular orbital (HOMO) calculated by SCAN, PBE0, and SCAN(SIC), respectively, for the NaCl molecule with bond lengths of $2.5$ \AA, $5.5$ \AA, $6.7$ \AA, and $7.3$ \AA  in Figure~\ref{Fig-NaCl-cube}. 
In the equilibrium region around $2.5$ \AA, the HOMO calculated by all three methods is primarily composed of the Cl $3p$ orbital. 
However, as the bond length increases, the calculated HOMOs begin to diverge significantly between different methods. 
The evolution of HOMO calculated by SCAN(SIC) accurately captures the correct electron transfer along the 
dissociation curve, resulting in a HOMO that is entirely composed of the Na $3s$ orbital at a bond length of 7.3 \AA.
Figure~\ref{Fig:Li-F-xDH} shows that semi-local SCAN and hybrid PBE0 predict inaccurate dissociation limits, 
characterized by a similar fractional charge distribution. 
Interestingly, our results indicate that SCAN tends to maintain a HOMO on the Cl $3p$ orbital, while the Na $3s$ orbital 
becomes more prominent in PBE0's HOMO at larger bond lengths. 
It is noteworthy that the xDH@SCAN curves deviate from the correct dissociation limit in a direction opposite to that of xDH@PBE0. 
This discrepancy may be attributed to the subtle differences in the HOMOs calculated by the two methods.

In a manner similar to the SIE4x4 dataset, we have compiled a dataset for heterodimers, 
which we refer to as HD2x5. 
This dataset contains 5 data points along the dissociation curves of NaCl and LiF, respectively.  
For LiF, the points are sampled at $1.5$ \AA, $2.3$ \AA, $3.1$ \AA, $4.7$ \AA, and $6.0$ \AA, while for NaCl, the points are sampled at $2.6$ \AA, $3.8$ \AA, $4.0$ \AA, $5.6$ \AA, and $7.8$ \AA. 
Using the CCSD(T) results as the reference, the MAEs of different methods are summarized in Table~\ref{table-mae}. 
Although using HF density impairs the performance of SCAN on the HD2x5 dataset, we find that a more accurate density is indeed beneficial.
This is evidenced by the superior performance of SCAN(SIC) and xDH@SCAN(SIC), which yield MAEs of only 2.64 kcal/mol and 2.24 kcal/mol, respectively.

Finally we show the impact of reference density and orbitals on more practical systems, 
which include 8 decomposition reactions of alkaline metal complexes in the ALK8 dataset, 
55 atomization energies in the G2-1 dataset, 
and 140 atomization energies in the W4-11 dataset \cite{cite-G2-1,cite-G2-2,cite-GMTKN55-detail}. 
The MAEs of different methods are also collected in Table~\ref{table-mae}.
We find that merely replacing the self-consistent SCAN density with the HF density does not enhance the overall performance of the semi-local SCAN functional. 
The MAEs of SCAN@HF exceed those of SCAN by 1.63 kcal/mol and 2.82 kcal/mol for the G2-1 and W4-11 datasets, respectively.
It is noted that the HF density generally underperforms compared to the self-consistent density of SCAN, 
a fact that is corroborated by the charge analysis on the dissociation curves of heterodimers mentioned above.   
Recently, the DC(HF)-DFT method was proposed \cite{dc-improved}, which advocates the use of HF density only if the density-driven error is significant.
Our observation suggests that the PZ-SIC scheme substantially enhances the SCAN density and, consequently, the SCAN results.  
The performance of SCAN(SIC) is distinctly superior to that of SCAN.
We have also examined the impact of the five empirical parameters by analyzing the performance of xDH[SCAN]@SCAN(SIC), which applies the parameters of xDH@SCAN with the SCAN(SIC) density. 
When comparing the performance of xDH[SCAN]@SCAN(SIC) with xDH@SCAN(SIC), both of which utilize the self-consistent SCAN(SIC) density, there is a clear and consistent improvement when moving from the former to the latter. 
This underlines the significant impact of fine-tuning the five parameters to enhance the precision of different xDH approaches (Please see Supplementary Information for a detailed discussion).
As illustrated in Table~\ref{table-mae} and Figure~\ref{Fig:mae}, the XYG3-type doubly hybrid method of xDH@SCAN(SIC) emerges as the overall winner when considering all six datasets, exhibiting minimal functional-driven and density-driven errors.

\section{Conclusion} \label{sec:conslusion}
It is widely recognized that two primary sources of errors exist in DFAs:
density-driven errors and functional-driven errors. 
This study demonstrated that the PZ-SIC scheme can effectively mitigate the density errors exhibited by semi-local and hybrid DFAs. This, in turn, reduces the corresponding density-driven errors in the density-sensitive scenarios, such as those observed in the BH76 and HD2x5 datasets. 
Importantly, this improvement can be achieved without compromising the density for the majority of systems that are density insensitive.

We demonstrate that xDHs have smaller functional-driven errors and are less susceptible to density errors compared to semi-local and hybrid DFAs.
By utilizing the semi-local SCAN exchange and correlation and the self-interaction corrected SCAN density, 
we have proposed a 5-parameter xDH method, xDH@SCAN(SIC). This method effectively minimizes both density-driven errors and functional-driven errors, 
thereby providing a balanced and accurate description across various challenging systems.

\section*{supplementary Information}
The following file is available free of charge:
\begin{itemize}
	\item supplementary.pdf, containing additional information needed for detailed discussions.
\end{itemize}
\begin{acknowledgments}
	This work was supported by the National Natural Science Foundation of China (21973015, 22125301, 22233002), 
	the Innovation Program for Quantum Science and Technology (2021ZD0303305), the Science Challenge Project (TZ2018004)
	Innovative Research Team of High-Level Local universities in Shanghai, and a Key Laboratory Program of the Education 
	Commission of Shanghai Municipality (ZDSYS14005).
\end{acknowledgments}

\bibliographystyle{apsrev4-1}
\bibliography{apssamp}
\clearpage 
\newpage
\begin{appendix}

    \section{Computed Results for Tests Set} \label{appSec:G2-1}
    The calculations in this work were all performed using the \ti{FHI-aims} package \cite{cite-aims-1} in the numerical integration framework. 
    We employed the numerical atom-centered orbital (NAO) basis sets with the \ti{tight} numerical integration setting for the atoms and molecules investigated in this work.

    The molecular geometries and the relevant reference values of the databases W4-11, BH76, ALK8 and SIE4x4 (Table.~\ref{tb:G2-1-DFA-formation-energy}-\ref{tb:HD2x5}) are collected online from the reference \cite{cite-GMTKN55-detail}.
    The molecular geometries and the relevant reference values of the G2-1 database are taken from references \cite{cite-G2-1,cite-G2-2} and references therein. 
    The Python scripts for the geometry transformation to the \ti{FHI-aims} format were carried out from the reference \cite{xDH-GMTKN55}.

    The HD2x5 database is compiled for heterodimers, which contains 5 data points along the dissociation curves of NaCl and LiF, respectively. 
    For LiF, the data points are sampled at $1.5$ \AA, $2.3$ \AA, $3.1$ \AA, $4.7$ \AA, and $6.0$ \AA, while for NaCl, 
    the data points are sampled at $2.6$ \AA, $3.8$ \AA, $4.0$ \AA, $5.6$ \AA, and $7.8$ \AA. 
    The CCSD(T) results are used as the reference energies for this database.
    The Hirshfeld charges of LiF and NaCl computed at the CCSD level are used as the reference for the charge analyses, which are obtained using the \ti{Gaussian} package at the basis set level of cc-pVTZ.

    \LTcapwidth=\textwidth
    \newcommand{\SCANHF}{     \begin{tabular}[c]{@{}c@{}}SCAN\\ @HF\end{tabular}}
    \newcommand{\SCANSIC}{    \begin{tabular}[c]{@{}c@{}}SCAN\\ (SIC)\end{tabular}}
    \newcommand{\xDHPBE}{   \begin{tabular}[c]{@{}c@{}}xDH\\ @PBE0 \end{tabular}}
    \newcommand{\xDHSCAN}{   \begin{tabular}[c]{@{}c@{}}xDH\\ @SCAN \end{tabular}}
    \newcommand{\xDHSCANSIC}{ \begin{tabular}[c]{@{}c@{}}xDH\\@SCAN\\(SIC)\end{tabular}}

          \begin{longtable*}[ht]{c|rrrrrrr}
        \caption{Atomization energy (in kcal/mol) 
        with the experimental reference values (Ref.) \cite{cite-G2-1,cite-G2-2} and the calculated results of the SCAN, SCAN@HF, SCAN(SIC), xDH@PBE0, xDH@SCAN and xDH@SCAN(SIC) methods 
        for the G2-1 database of 55 small molecules.
        }
        \label{tb:G2-1-DFA-formation-energy} \\
      \hline\hline
      Systems        & Expt.     & SCAN    & \SCANHF & \SCANSIC & \xDHPBE & \xDHSCAN & \xDHSCANSIC \\ \hline
      \ce{LiH      } & -57.98    & -56.10  & -55.72  & -55.72   & -59.14   & -57.33   & -59.61        \\ 
      \ce{BeH      } & -49.92    & -60.79  & -59.61  & -60.36   & -53.50   & -54.86   & -55.11        \\ 
      \ce{CH       } & -83.99    & -82.20  & -81.08  & -81.81   & -85.15   & -81.33   & -83.06        \\ 
      \ce{CH2 }(1A1) & -190.37   & -197.57 & -196.29 & -196.83  & -191.23  & -192.75  & -193.45       \\ 
      \ce{CH2 }(3B1) & -180.92   & -175.85 & -174.07 & -175.22  & -181.70  & -175.90  & -178.70       \\ 
      \ce{CH3      } & -307.55   & -313.42 & -311.49 & -312.48  & -309.59  & -307.79  & -309.83       \\ 
      \ce{CH4      } & -420.19   & -420.65 & -418.12 & -419.53  & -422.19  & -417.64  & -420.30       \\ 
      \ce{NH       } & -83.58    & -85.87  & -84.27  & -85.22   & -85.08   & -81.54   & -83.51        \\ 
      \ce{NH2      } & -181.72   & -185.89 & -183.61 & -184.80  & -185.27  & -179.33  & -182.72       \\ 
      \ce{NH3      } & -297.93   & -296.38 & -293.76 & -295.09  & -300.19  & -292.38  & -296.15       \\ 
      \ce{OH       } & -106.71   & -108.82 & -107.85 & -108.30  & -106.81  & -104.13  & -105.87       \\ 
      \ce{H2O      } & -232.78   & -229.79 & -228.10 & -228.96  & -231.39  & -226.78  & -229.06       \\ 
      \ce{HF       } & -141.42   & -136.98 & -135.96 & -136.39  & -140.10  & -137.55  & -138.53       \\ 
      \ce{SiH2}(1A1) & -151.91   & -150.46 & -147.54 & -149.03  & -155.17  & -153.09  & -155.52       \\ 
      \ce{SiH2}(3B1) & -131.16   & -139.26 & -136.33 & -138.38  & -133.80  & -136.84  & -137.59       \\ 
      \ce{SiH3     } & -225.84   & -231.82 & -227.90 & -230.40  & -230.36  & -231.61  & -233.42       \\ 
      \ce{SiH4     } & -322.39   & -324.93 & -320.36 & -323.31  & -327.59  & -327.05  & -329.92       \\ 
      \ce{PH2      } & -152.93   & -157.94 & -153.97 & -156.29  & -157.37  & -154.33  & -156.78       \\ 
      \ce{PH3      } & -241.92   & -243.35 & -238.15 & -241.08  & -245.23  & -240.75  & -243.84       \\ 
      \ce{SH2      } & -182.98   & -183.41 & -180.70 & -181.84  & -184.53  & -182.17  & -183.65       \\ 
      \ce{HCl      } & -107.22   & -106.15 & -104.86 & -104.87  & -107.19  & -106.31  & -106.70       \\ 
      \ce{Li2      } & -24.37    & -18.47  & -18.59  & -18.38   & -26.67   & -25.34   & -27.99        \\ 
      \ce{LiF      } & -137.87   & -134.79 & -131.94 & -133.27  & -140.27  & -139.03  & -141.00       \\ 
      \ce{C2H2     } & -405.79   & -403.96 & -398.55 & -401.53  & -408.97  & -406.38  & -409.10       \\ 
      \ce{C2H4     } & -563.55   & -564.88 & -559.54 & -562.72  & -567.10  & -561.30  & -564.93       \\ 
      \ce{C2H6     } & -712.27   & -715.23 & -709.98 & -712.78  & -717.30  & -710.55  & -714.42       \\ 
      \ce{CN       } & -181.51   & -179.91 & -161.82 & -176.05  & -187.84  & -181.52  & -185.26       \\ 
      \ce{HCN      } & -312.76   & -310.24 & -303.97 & -307.84  & -318.74  & -311.62  & -315.75       \\ 
      \ce{CO       } & -259.61   & -255.18 & -250.42 & -254.42  & -260.90  & -254.55  & -257.10       \\ 
      \ce{HCO      } & -279.03   & -283.33 & -277.20 & -281.66  & -281.17  & -276.60  & -279.53       \\ 
      \ce{H2CO     } & -374.01   & -374.86 & -368.67 & -372.62  & -376.68  & -369.64  & -373.62       \\ 
      \ce{CH3OH    } & -512.81   & -514.31 & -508.69 & -512.28  & -514.87  & -507.41  & -511.58       \\ 
      \ce{N2       } & -228.51   & -222.00 & -216.17 & -220.58  & -234.65  & -221.05  & -226.15       \\ 
      \ce{N2H4     } & -438.33   & -438.19 & -432.08 & -435.60  & -443.69  & -430.22  & -436.96       \\ 
      \ce{NO       } & -152.95   & -152.58 & -144.71 & -151.43  & -155.68  & -147.20  & -151.28       \\ 
      \ce{O2       } & -120.73   & -127.84 & -121.06 & -126.49  & -114.89  & -115.17  & -115.80       \\ 
      \ce{H2O2     } & -269.14   & -268.19 & -262.10 & -266.89  & -267.52  & -257.83  & -264.09       \\ 
      \ce{F2       } & -39.09    & -36.56  & -30.17  & -33.38   & -37.84   & -29.89   & -35.04        \\ 
      \ce{CO2      } & -389.7    & -394.81 & -384.15 & -391.19  & -393.25  & -387.16  & -391.37       \\ 
      \ce{Na2      } & -17.03    & -14.06  & -14.08  & -13.94   & -25.17   & -22.31   & -23.04        \\ 
      \ce{Si2      } & -75.57    & -77.98  & -74.30  & -77.12   & -72.37   & -75.24   & -74.22        \\ 
      \ce{P2       } & -117.22   & -114.71 & -107.03 & -111.72  & -122.85  & -119.42  & -121.91       \\ 
      \ce{S2       } & -102.71   & -110.52 & -104.38 & -107.76  & -102.43  & -104.61  & -103.98       \\ 
      \ce{Cl2      } & -59.59    & -58.28  & -53.62  & -56.45   & -58.81   & -57.56   & -58.22        \\ 
      \ce{NaCl     } & -98.84    & -97.52  & -96.71  & -96.41   & -100.14  & -100.86  & -101.09       \\ 
      \ce{SiO      } & -192.91   & -188.87 & -181.96 & -186.57  & -194.12  & -193.84  & -196.54       \\ 
      \ce{CS       } & -172.01   & -168.41 & -162.52 & -165.24  & -174.70  & -172.29  & -173.76       \\ 
      \ce{SO       } & -125.81   & -132.21 & -124.87 & -130.54  & -123.63  & -124.19  & -124.53       \\ 
      \ce{ClO      } & -65.58    & -69.37  & -58.75  & -65.93   & -63.95   & -61.55   & -63.52        \\ 
      \ce{ClF      } & -62.55    & -60.70  & -54.60  & -59.07   & -61.58   & -57.42   & -59.68        \\ 
      \ce{Si2H6    } & -531.27   & -538.82 & -529.63 & -535.40  & -541.27  & -542.23  & -546.27       \\ 
      \ce{CH3Cl    } & -395.39   & -397.42 & -392.83 & -394.96  & -398.92  & -395.22  & -397.23       \\ 
      \ce{CH3SH    } & -474.25   & -477.10 & -471.38 & -474.00  & -478.61  & -473.69  & -476.65       \\ 
      \ce{HOCl     } & -165.43   & -164.83 & -159.09 & -163.38  & -165.43  & -159.74  & -163.13       \\ 
      \ce{SO2      } & -259.16   & -262.47 & -247.02 & -256.47  & -263.16  & -259.64  & -264.30       \\ \hline\hline
      \end{longtable*}
    
    \begin{longtable*}[ht]{c|rrrrrrr}
      \caption{
      Atomization energy (in kcal/mol) 
      with the highly accurate Weizmann-4 results as references (Ref.) and the calculated results of the SCAN, SCAN@HF, SCAN(SIC), xDH@PBE0, xDH@SCAN and xDH@SCAN(SIC) methods
      for the W4-11 database of 99 small molecules and 41 medium-sized organic molecules.
      }
      \label{tb:W4-11-DFA-formation-energy} \\
      \hline\hline
      Systems      & Ref.    & SCAN    & \SCANHF & \SCANSIC & \xDHPBE & \xDHSCAN & \xDHSCANSIC \\ \hline
      \ce{H2}           & 109.49  & 107.67  & 107.00  & 107.71   & 109.61   & 109.04   & 107.04       \\ 
      \ce{AlH3}         & 213.17  & 212.46  & 210.85  & 213.44   & 215.34   & 215.56   & 216.12       \\
      \ce{AlH}          & 73.57   & 69.81   & 69.52   & 70.36    & 73.88    & 72.92    & 74.27        \\
      \ce{SiH4}         & 324.95  & 324.01  & 319.90  & 324.58   & 328.31   & 328.36   & 330.13       \\
      \ce{BH3}          & 281.29  & 283.07  & 281.14  & 283.16   & 283.85   & 283.23   & 281.21       \\
      \ce{BH}           & 85.00   & 79.37   & 78.41   & 79.35    & 85.18    & 83.76    & 83.39        \\
      \ce{CH2}-trip     & 190.75  & 197.08  & 196.18  & 197.42   & 191.37   & 193.36   & 192.94       \\
      \ce{SiH}          & 73.92   & 72.52   & 71.25   & 72.83    & 75.01    & 73.88    & 75.18        \\
      \ce{Si2H6}        & 535.89  & 537.05  & 529.31  & 537.83   & 541.76   & 544.65   & 547.30       \\
      \ce{CH3}          & 307.87  & 312.76  & 311.28  & 313.19   & 310.20   & 310.07   & 309.02       \\
      \ce{CH4}          & 420.42  & 419.92  & 417.69  & 420.49   & 423.07   & 421.25   & 419.45       \\
      \ce{B2H6}         & 607.02  & 617.23  & 611.36  & 617.38   & 610.69   & 611.81   & 609.45       \\
      \ce{SiH3F}        & 382.75  & 379.10  & 374.20  & 379.43   & 382.91   & 383.47   & 388.24       \\
      \ce{PH3}          & 242.27  & 241.83  & 238.71  & 243.25   & 246.04   & 243.71   & 244.52       \\
      \ce{C2H6}         & 713.08  & 713.73  & 709.74  & 714.87   & 718.08   & 715.02   & 713.19       \\
      Propane      & 1007.91 & 1009.58 & 1003.66 & 1008.61  & 1014.90  & 1010.68  & 1009.59      \\
      \ce{CH2}-sing     & 181.46  & 175.42  & 173.89  & 175.63   & 182.49   & 178.89   & 178.33       \\
      \ce{CH}           & 84.22   & 81.94   & 81.00   & 81.98    & 85.63    & 83.05    & 82.70        \\
      \ce{H2S}          & 183.91  & 182.45  & 181.08  & 183.26   & 184.47   & 183.82   & 184.01       \\
      \ce{HS}           & 87.73   & 88.84   & 88.01   & 89.47    & 88.41    & 87.81    & 88.60        \\
      \ce{C2H5F}        & 721.50  & 722.70  & 716.24  & 723.88   & 723.74   & 719.53   & 721.55       \\
      \ce{CH3NH2}       & 582.30  & 581.70  & 577.70  & 582.72   & 588.19   & 580.49   & 581.31       \\
      \ce{CH3F}         & 422.96  & 422.53  & 418.08  & 423.18   & 423.26   & 420.09   & 421.42       \\
      Propene      & 861.58  & 862.72  & 857.13  & 861.56   & 867.06   & 862.53   & 862.87       \\
      \ce{NH3}          & 298.02  & 295.46  & 293.70  & 296.06   & 301.37   & 295.14   & 295.30       \\
      Ethanol      & 811.24  & 812.07  & 805.75  & 810.05   & 814.45   & 808.99   & 809.67       \\
      \ce{CH3NH}        & 474.63  & 480.28  & 475.55  & 481.00   & 480.35   & 473.90   & 475.85       \\
      \ce{C2H4}         & 564.10  & 563.31  & 559.75  & 564.36   & 567.52   & 564.11   & 563.95       \\
      Methanol     & 513.50  & 512.94  & 508.61  & 513.74   & 514.75   & 510.35   & 510.60       \\
      \ce{HCl}          & 107.50  & 105.57  & 105.10  & 106.25   & 106.79   & 106.75   & 106.98       \\
      \ce{NH2}          & 182.59  & 185.07  & 183.72  & 185.62   & 186.16   & 181.15   & 182.04       \\
      \ce{NH}           & 83.10   & 85.31   & 84.50   & 85.70    & 85.58    & 82.17    & 83.17        \\
      \ce{CH2NH2}       & 482.28  & 487.35  & 482.90  & 486.01   & 487.33   & 481.32   & 482.33       \\
      \ce{BHF2}         & 410.97  & 410.88  & 405.92  & 409.66   & 406.39   & 406.76   & 412.73       \\
      \ce{H2O}          & 232.97  & 229.19  & 227.85  & 229.54   & 231.51   & 228.82   & 228.38       \\
      \ce{HF}           & 141.64  & 136.58  & 136.00  & 136.86   & 139.02   & 137.90   & 138.20       \\
      \ce{CH2CH}        & 446.08  & 451.23  & 445.00  & 451.90   & 448.53   & 446.69   & 447.95       \\
      \ce{OH}           & 107.21  & 108.51  & 107.70  & 108.61   & 106.94   & 105.35   & 105.42       \\
      Propyne      & 705.61  & 704.83  & 700.42  & 706.63   & 710.92   & 706.35   & 710.54       \\
      Acetaldehyde & 677.86  & 679.64  & 672.68  & 680.26   & 681.36   & 675.70   & 677.85       \\
      Allene       & 704.10  & 707.05  & 701.57  & 708.10   & 708.08   & 703.04   & 706.77       \\
      \ce{SiF4}         & 577.78  & 565.87  & 558.12  & 564.61   & 564.96   & 568.28   & 580.37       \\
      \ce{BF3}          & 470.97  & 469.17  & 464.01  & 469.87   & 462.60   & 463.65   & 475.02       \\
      \ce{C2H3F}        & 573.89  & 575.24  & 568.79  & 572.81   & 574.46   & 570.08   & 573.50       \\
      Oxirane      & 651.53  & 655.08  & 648.24  & 652.65   & 655.28   & 649.29   & 651.17       \\
      \ce{CH2F2}        & 437.67  & 437.26  & 430.23  & 436.39   & 435.24   & 431.25   & 435.03       \\
      \ce{AlF3}         & 430.97  & 421.91  & 416.59  & 421.46   & 420.49   & 423.22   & 433.67       \\
      \ce{BeF2}         & 309.10  & 310.11  & 307.94  & 311.17   & 303.32   & 305.41   & 312.76       \\
      \ce{CH2C}         & 359.93  & 357.86  & 353.36  & 358.39   & 361.62   & 356.83   & 358.27       \\
      \ce{N2H4}         & 438.28  & 436.19  & 432.45  & 437.57   & 445.00   & 432.45   & 435.85       \\
      \ce{CH2NH}        & 439.44  & 437.67  & 433.60  & 438.74   & 444.58   & 436.53   & 438.20       \\
      \ce{AlF}          & 163.78  & 158.36  & 156.35  & 158.24   & 159.55   & 159.79   & 163.44       \\
      Acetic       & 804.02  & 807.88  & 797.89  & 803.65   & 805.66   & 798.64   & 802.73       \\
      \ce{C2H2}         & 405.53  & 401.96  & 399.53  & 403.54   & 408.55   & 405.24   & 408.32       \\
      \ce{H2CO}         & 374.66  & 373.56  & 368.89  & 373.68   & 376.40   & 371.78   & 372.45       \\
      \ce{H2CN}         & 343.75  & 347.38  & 339.24  & 348.21   & 348.11   & 341.67   & 344.72       \\
      \ce{BF}           & 182.52  & 179.57  & 176.38  & 179.61   & 179.40   & 178.42   & 182.06       \\
      \ce{BeCl2}        & 225.27  & 233.06  & 233.06  & 234.36   & 223.85   & 229.47   & 234.32       \\
      t-\ce{HCOH}       & 322.48  & 319.88  & 314.14  & 320.05   & 322.80   & 317.25   & 318.27       \\
      \ce{AlCl3}        & 312.65  & 313.33  & 312.73  & 312.45   & 308.32   & 317.15   & 321.18       \\
      c-\ce{HCOH}       & 317.65  & 315.18  & 309.02  & 315.44   & 317.87   & 312.26   & 312.71       \\
      \ce{AlCl}         & 122.62  & 120.08  & 119.77  & 120.79   & 120.22   & 122.08   & 124.31       \\
      Ketene       & 533.46  & 538.19  & 531.09  & 537.29   & 536.51   & 531.02   & 533.74       \\
      \ce{SiF}          & 142.71  & 140.00  & 136.53  & 138.69   & 139.92   & 139.42   & 142.85       \\
      Formic       & 501.90  & 503.93  & 495.66  & 503.71   & 501.97   & 495.97   & 499.90       \\
      \ce{HCNH}         & 336.25  & 339.36  & 332.38  & 339.78   & 340.99   & 334.58   & 336.65       \\
      Glyoxal      & 635.10  & 636.19  & 626.73  & 636.52   & 637.50   & 629.25   & 635.30       \\
      \ce{HCOF}         & 403.74  & 405.03  & 396.89  & 404.49   & 402.77   & 397.83   & 402.56       \\
      \ce{NH2Cl}        & 248.06  & 246.47  & 243.04  & 247.78   & 250.64   & 244.44   & 247.38       \\
      \ce{CF4}          & 478.76  & 478.64  & 467.50  & 478.65   & 470.20   & 465.15   & 475.95       \\
      \ce{HCCF}         & 398.47  & 397.37  & 392.20  & 397.91   & 398.93   & 394.60   & 401.07       \\
      \ce{HCN}          & 313.42  & 308.25  & 305.03  & 309.79   & 318.94   & 311.34   & 314.86       \\
      \ce{HNC}          & 298.20  & 294.24  & 291.16  & 295.54   & 302.61   & 294.76   & 297.02       \\
      \ce{CCH}          & 266.16  & 266.44  & 259.71  & 267.54   & 266.92   & 265.49   & 268.74       \\
      \ce{HCO}          & 279.42  & 282.33  & 277.45  & 282.77   & 280.66   & 277.69   & 278.91       \\
      \ce{CO}           & 259.73  & 254.46  & 250.64  & 254.97   & 260.60   & 256.11   & 256.77       \\
      Oxirene      & 456.07  & 456.52  & 450.64  & 454.17   & 458.03   & 451.97   & 457.85       \\
      \ce{F2CO}         & 420.64  & 423.26  & 412.95  & 421.99   & 416.85   & 411.14   & 418.80       \\
      \ce{HOCN}         & 410.07  & 408.20  & 402.05  & 409.15   & 414.34   & 404.52   & 410.75       \\
      \ce{HOOH}         & 269.09  & 267.15  & 262.18  & 267.67   & 267.00   & 260.00   & 263.15       \\
      t-\ce{N2H2}       & 296.53  & 291.99  & 287.09  & 291.86   & 303.07   & 289.88   & 290.93       \\
      \ce{HNCO}         & 434.74  & 439.15  & 431.47  & 437.76   & 440.19   & 431.20   & 434.56       \\
      c-\ce{N2H2}       & 291.14  & 286.61  & 281.44  & 287.57   & 297.68   & 284.33   & 287.92       \\
      \ce{CF2}          & 258.78  & 257.30  & 247.93  & 254.40   & 254.86   & 250.31   & 254.16       \\
      \ce{CO2}          & 390.14  & 393.11  & 385.04  & 392.02   & 391.86   & 386.16   & 389.30       \\
      \ce{FCCF}         & 386.09  & 386.09  & 378.86  & 386.53   & 383.69   & 378.37   & 388.06       \\
      Dioxirane    & 410.03  & 412.96  & 403.57  & 412.96   & 412.16   & 404.33   & 408.31       \\
      \ce{CF}           & 132.72  & 133.77  & 128.21  & 132.83   & 131.74   & 129.02   & 130.70       \\
      \ce{SHS}          & 165.13  & 168.80  & 164.15  & 169.73   & 164.66   & 165.62   & 167.31       \\
      \ce{HOCl}         & 166.23  & 164.24  & 160.00  & 164.88   & 164.35   & 161.25   & 163.55       \\
      \ce{NCCN}         & 502.04  & 498.17  & 490.15  & 499.56   & 513.32   & 499.46   & 509.29       \\
      \ce{N2}           & 228.49  & 220.68  & 216.86  & 221.77   & 236.11   & 223.64   & 225.31       \\
      \ce{N2H}          & 224.86  & 225.92  & 216.80  & 225.76   & 231.45   & 220.26   & 222.64       \\
      \ce{OCS}          & 335.75  & 340.41  & 333.30  & 338.82   & 338.40   & 335.63   & 336.94       \\
      \ce{SiO}          & 193.05  & 187.87  & 182.65  & 187.82   & 193.33   & 193.24   & 196.43       \\
      \ce{ClCN}         & 285.45  & 283.11  & 277.42  & 283.93   & 290.13   & 283.61   & 288.56       \\
      \ce{HOO}          & 175.53  & 179.63  & 170.46  & 178.24   & 173.53   & 168.42   & 170.24       \\
      \ce{HCNO}         & 364.97  & 368.89  & 358.76  & 366.22   & 372.00   & 363.21   & 368.81       \\
      \ce{HONC}         & 350.15  & 347.97  & 340.31  & 347.87   & 353.71   & 342.40   & 348.29       \\
      \ce{HNO}          & 205.89  & 200.75  & 195.19  & 201.20   & 209.15   & 199.14   & 201.16       \\
      \ce{HOF}          & 158.65  & 156.82  & 151.83  & 157.00   & 155.99   & 150.04   & 154.62       \\
      c-\ce{HONO}       & 312.22  & 313.64  & 301.06  & 310.91   & 314.93   & 303.48   & 307.82       \\
      t-\ce{HONO}       & 312.65  & 314.49  & 301.72  & 311.57   & 315.41   & 303.95   & 308.03       \\
      \ce{CS2}          & 280.78  & 286.37  & 280.98  & 279.34   & 283.72   & 284.59   & 279.34       \\
      \ce{HNNN}         & 331.79  & 335.51  & 325.01  & 333.53   & 345.32   & 328.59   & 333.30       \\
      \ce{CS}           & 172.22  & 167.24  & 163.50  & 166.28   & 173.94   & 172.82   & 171.73       \\
      \ce{CN}           & 181.35  & 178.55  & 162.42  & 176.90   & 187.89   & 182.95   & 181.16       \\
      \ce{SO3}          & 346.94  & 352.13  & 337.37  & 349.23   & 346.07   & 342.94   & 349.81       \\
      \ce{CCl2}         & 177.36  & 179.47  & 170.20  & 176.82   & 175.44   & 174.99   & 177.39       \\
      \ce{BN3PI}        & 105.82  & 114.60  & 110.30  & 114.76   & 107.81   & 105.91   & 108.96       \\
      \ce{SO2}          & 260.62  & 260.74  & 250.44  & 257.77   & 259.66   & 258.49   & 262.23       \\
      \ce{NO}           & 152.75  & 151.57  & 145.03  & 152.08   & 156.28   & 148.21   & 150.70       \\
      \ce{SO}           & 126.47  & 131.66  & 126.29  & 131.10   & 122.66   & 123.24   & 124.30       \\
      \ce{N2O}          & 270.85  & 273.71  & 262.32  & 270.35   & 281.59   & 268.14   & 271.99       \\
      c-\ce{HOOO}       & 233.09  & 241.62  & 218.59  & 236.89   & 230.00   & 223.68   & 225.91       \\
      \ce{S2}           & 104.25  & 109.26  & 106.60  & 109.41   & 100.96   & 103.38   & 104.20       \\
      \ce{P4}           & 290.58  & 292.88  & 287.60  & 297.29   & 302.91   & 302.88   & 305.92       \\
      \ce{Cl2}          & 59.75   & 57.93   & 55.41   & 58.61    & 57.19    & 57.34    & 60.21        \\
      \ce{O2}           & 120.82  & 127.19  & 121.41  & 127.24   & 114.48   & 113.52   & 115.05       \\
      \ce{F2}           & 39.04   & 35.60   & 30.84   & 34.10    & 35.76    & 30.56    & 30.92        \\
      t-\ce{HOOO}        & 233.30  & 242.38  & 221.22  & 234.53   & 230.67   & 225.63   & 221.16       \\
      \ce{S2O}          & 208.78  & 211.10  & 201.58  & 206.62   & 208.92   & 211.21   & 210.47       \\
      \ce{P2}           & 117.59  & 112.01  & 109.37  & 114.13   & 122.51   & 120.13   & 122.40       \\
      \ce{ClF}          & 62.80   & 60.51   & 56.33   & 60.51    & 59.58    & 57.39    & 61.45        \\
      \ce{NO2}          & 227.88  & 236.41  & 224.04  & 229.42   & 233.14   & 225.00   & 221.77       \\
      \ce{ClO}          & 65.45   & 69.34   & 60.16   & 67.43    & 62.94    & 61.86    & 62.81        \\
      \ce{S3}           & 168.36  & 171.05  & 163.96  & 168.83   & 168.62   & 174.29   & 173.56       \\
      \ce{Cl2O}         & 101.46  & 101.56  & 92.76   & 98.43    & 98.36    & 95.16    & 99.86        \\
      \ce{S4}-c2v       & 234.35  & 240.61  & 232.48  & 233.85   & 236.22   & 250.58   & 231.66       \\
      \ce{OF}           & 53.08   & 56.98   & 48.07   & 55.66    & 51.04    & 47.00    & 49.56        \\
      \ce{C2}           & 147.02  & 120.99  & 116.00  & 138.37   & 160.97   & 176.58   & 137.57       \\
      \ce{OCLO}         & 128.12  & 136.65  & 122.99  & 132.93   & 125.76   & 126.07   & 127.96       \\
      \ce{F2O}          & 93.78   & 95.99   & 84.95   & 95.36    & 89.52    & 80.83    & 89.70        \\
      \ce{B2}           & 67.46   & 69.83   & 66.63   & 76.27    & 65.69    & 75.26    & 60.96        \\
      \ce{FO2}          & 134.72  & 143.82  & 122.14  & 130.15   & 129.12   & 131.29   & 118.11       \\
      \ce{CLOO}         & 126.39  & 136.33  & 115.27  & 125.66   & 116.65   & 118.22   & 112.36       \\
      \ce{FOOF}         & 152.37  & 164.49  & 138.48  & 139.99   & 152.08   & 145.44   & 126.73       \\
      \ce{O3}           & 147.43  & 147.85  & 132.63  & 140.84   & 156.55   & 156.45   & 126.90       \\
      \ce{BN}           & 105.24  & 91.84   & 85.02   & 102.36   & 121.53   & 130.53   & 98.94        \\ \hline\hline
      \ce{Be2}          & 2.67    & 8.61    & 6.23    & 11.39    & 2.71     & 4.00     & 2.11        
    \end{longtable*}

          \begin{longtable*}[ht]{c|rrrrrrrr}
        \caption{
        Barrier heights (in kcal/mol) 
        with the Weizmann-1 results as references (Ref.) and the calculated results of the SCAN, SCAN@HF, SCAN(SIC), xDH@PBE0, xDH@SCAN and xDH@SCAN(SIC) methods. 
        for the 76 forward and reverse barrier heights in the BH76 test set \cite{cite-bh76-1, cite-bh76-2, cite-GMTKN55}.
        $V_f$ represents the forward reaction barrier height, and $V_b$ is the backward reaction barrier height. 
        }
        \label{tb:reaction} \\
        \hline\hline
          Reaction                                   &       & Ref.  & SCAN  & \SCANHF & SIC-SCAN & \xDHPBE & \xDHSCAN & \xDHSCANSIC\\ \hline
          \ce{H    + N2O   ->   OH    +  N2   }      & $V_f$ & 17.70 & 9.37  & 17.45  & 14.68  & 21.67  & 20.30  & 27.22  \\ 
                                                     & $V_b$ & 82.60 & 64.50 & 80.08  & 74.54  & 79.25  & 78.73  & 88.37  \\ 
          \ce{H    + HF    ->   HF    +  H    }      & $V_f$ & 42.10 & 29.19 & 32.11  & 34.39  & 44.89  & 43.67  & 46.99  \\ 
                                                     & $V_b$ & 42.10 & 29.19 & 32.11  & 34.39  & 44.89  & 43.67  & 46.99  \\ 
          \ce{H    + HCl   ->   HCl   +  H    }      & $V_f$ & 17.80 & 9.23  & 11.65  & 10.68  & 18.52  & 17.13  & 17.52  \\ 
                                                     & $V_b$ & 17.80 & 9.23  & 11.65  & 10.68  & 18.52  & 17.13  & 17.52  \\ 
          \ce{H    + FCH3  ->   HF    +  CH3  }      & $V_f$ & 30.50 & 20.27 & 25.27  & 25.10  & 33.84  & 32.94  & 35.95  \\ 
                                                     & $V_b$ & 56.90 & 47.00 & 54.75  & 51.97  & 58.28  & 59.76  & 61.84  \\ 
          \ce{H    + F2    ->   HF    +  F    }      & $V_f$ & 1.50  & -11.21& 10.69  & -3.12  & 5.75   & 3.28   & 9.52   \\ 
                                                     & $V_b$ & 104.80& 89.13 & 116.30 & 97.28  & 108.16 & 111.05 & 111.08 \\ 
          \ce{CH3  + ClF   ->   CH3F  +  Cl   }      & $V_f$ & 7.10  & -3.71 & 9.00   & 2.08   & 8.29   & 8.60   & 11.05  \\ 
                                                     & $V_b$ & 59.80 & 45.15 & 49.06  & 51.40  & 60.88  & 60.90  & 62.79  \\ 
          \ce{F-   + CH3F  ->   FCH3  +  F-   }      & $V_f$ & -0.60 & -6.72 & -3.26  & -6.23  & -1.68  & -1.03  & -3.33  \\ 
                                                     & $V_b$ & -0.60 & -6.72 & -3.26  & -6.23  & -1.68  & -1.03  & -3.33  \\ 
          \ce{F-\bond{...}CH3F  -> FCH3\bond{...}F-} & $V_f$ & 13.40 & 8.23  & 13.17  & 8.56   & 12.12  & 12.96  & 10.97  \\ 
                                                     & $V_b$ & 13.40 & 8.23  & 13.17  & 8.56   & 12.12  & 12.96  & 10.97  \\ 
          \ce{Cl-  + CH3Cl ->   ClCH3 +  Cl-  }      & $V_f$ & 2.50  & -6.99 & 1.10   & -5.69  & 0.85   & 0.99   & -0.52  \\ 
                                                     & $V_b$ & 2.50  & -6.99 & 1.10   & -5.69  & 0.85   & 0.99   & -0.52  \\ 
          \ce{Cl-\bond{...}CH3Cl -> ClCH3\bond{...}Cl-}& $V_f$ & 13.50 & 6.21  & 12.89  & 7.04 & 12.68  & 12.90  & 11.40  \\ 
                                                     & $V_b$ & 13.50 & 6.21  & 12.89  & 7.04   & 12.68  & 12.90  & 11.40  \\ 
          \ce{F-   + CH3Cl ->   Cl-   +  FCH3 }      & $V_f$ & -12.30& -19.58& -17.36 & -18.00 & -12.84 & -12.71 & -14.63 \\ 
                                                     & $V_b$ & 19.80 & 12.22 & 21.88  & 14.57  & 16.87  & 16.76  & 15.35  \\ 
          \ce{F-\bond{...}CH3Cl -> FCH3\bond{...}Cl-}  & $V_f$ & 3.50  & -1.09 & 2.08   & -0.32& 2.86   & 2.97   & 1.85   \\ 
                                                     & $V_b$ & 29.60 & 23.30 & 31.82  & 25.50  & 27.60  & 27.80  & 26.16  \\ 
          \ce{OH-  + CH3F  ->   OHCH3 +  F-   }      & $V_f$ & -2.70 & -10.85& -4.68  & -6.29  & -5.36  & -5.31  & -5.45  \\ 
                                                     & $V_b$ & 17.60 & 11.32 & 14.39  & 15.34  & 16.54  & 17.47  & 16.26  \\ 
          \ce{OH-\bond{...}CH3F -> OHCH3\bond{...}F-}  & $V_f$ & 11.00 & 4.42  & 10.59  & 8.71 & 8.90   & 9.50   & 9.12   \\ 
                                                     & $V_b$ & 47.70 & 44.26 & 50.31  & 43.85  & 47.41  & 48.33  & 45.62  \\ 
          \ce{H    + N2    ->   HN2           }      & $V_f$ & 14.60 & 4.23  & 10.25  & 7.38   & 15.24  & 14.03  & 16.40  \\ 
                                                     & $V_b$ & 10.90 & 9.73  & 9.83   & 11.30  & 12.23  & 11.97  & 13.04  \\ 
          \ce{H    + CO    ->   HCO           }      & $V_f$ & 3.20  & -3.80 & -0.54  & -3.00  & 3.87   & 3.00   & 2.33   \\ 
                                                     & $V_b$ & 22.80 & 24.35 & 26.30  & 24.72  & 25.01  & 25.13  & 23.97  \\ 
          \ce{H    + C2H4  ->   CH3CH2        }      & $V_f$ & 2.00  & -4.54 & -1.32  & -3.85  & 3.07   & 2.02   & 2.36   \\ 
                                                     & $V_b$ & 42.00 & 43.10 & 46.44  & 43.55  & 44.12  & 45.03  & 44.73  \\ 
          \ce{CH3  + C2H4  ->   CH3CH2        }      & $V_f$ & 6.40  & 0.56  & 5.41   & 1.33   & 7.10   & 6.70   & 7.66   \\ 
                                                     & $V_b$ & 33.00 & 30.64 & 34.69  & 28.61  & 35.97  & 36.21  & 36.03  \\ 
          \ce{HCN          ->   HNC           }      & $V_f$ & 48.10 & 46.63 & 47.26  & 47.83  & 50.53  & 50.30  & 50.33  \\ 
                                                     & $V_b$ & 33.00 & 32.36 & 33.61  & 33.61  & 32.69  & 32.87  & 33.85  \\ 
          \ce{H    + HCl   ->   H2    +  Cl   }      & $V_f$ & 6.10  & -1.40 & 0.83   & -0.57  & 6.85   & 5.83   & 4.61   \\ 
                                                     & $V_b$ & 8.00  & 0.08  & -7.29  & 0.86   & 7.24   & 5.97   & 5.87   \\ 
          \ce{OH   + H2    ->   H2O   +  H    }      & $V_f$ & 5.20  & -2.18 & 4.22   & -0.08  & 5.58   & 5.36   & 5.35   \\ 
                                                     & $V_b$ & 21.60 & 11.16 & 16.83  & 13.14  & 21.83  & 21.39  & 20.51  \\ 
          \ce{CH3  + H2    ->   CH4   +  H    }      & $V_f$ & 11.90 & 7.28  & 9.98   & 7.81   & 11.27  & 11.90  & 11.52  \\ 
                                                     & $V_b$ & 15.00 & 6.87  & 9.00   & 7.37   & 15.58  & 15.15  & 14.32  \\ 
          \ce{OH   + CH4   ->   H2O   +  CH3  }      & $V_f$ & 6.30  & -1.89 & 4.33   & 0.33   & 6.22   & 6.40   & 6.98   \\ 
                                                     & $V_b$ & 19.50 & 11.86 & 17.92  & 13.98  & 18.16  & 19.18  & 19.33  \\ 
          \ce{H    + H2    ->   H     +  H2   }      & $V_f$ & 9.70  & 2.41  & 3.88   & 2.93   & 9.83   & 9.16   & 8.89   \\ 
                                                     & $V_b$ & 9.70  & 2.41  & 3.88   & 2.93   & 9.83   & 9.16   & 8.89   \\ 
          \ce{OH   + NH3   ->   H2O   +  NH2  }      & $V_f$ & 3.40  & -7.40 & 1.82   & -4.61  & 3.62   & 3.68   & 3.95   \\ 
                                                     & $V_b$ & 13.70 & 3.04  & 11.88  & 5.87   & 12.78  & 13.23  & 13.91  \\ 
          \ce{HCl  + CH3   ->   Cl    +  CH4  }      & $V_f$ & 1.80  & -2.93 & 0.19   & -2.24  & 1.45   & 2.20   & 1.60   \\ 
                                                     & $V_b$ & 6.80  & -1.86 & -8.92  & -1.25  & 6.16   & 5.60   & 5.66   \\ 
          \ce{OH   + C2H6  ->   H2O   +  C2H5 }      & $V_f$ & 3.50  & -5.08 & 2.61   & 1.00   & 3.62   & 3.76   & 4.21   \\ 
                                                     & $V_b$ & 20.40 & 13.14 & 20.09  & 18.92  & 18.97  & 20.04  & 20.16  \\ 
          \ce{F    + H2    ->   HF    +  H    }      & $V_f$ & 1.60  & -7.43 & 2.02   & -3.34  & 3.25   & 3.92   & 1.60   \\ 
                                                     & $V_b$ & 33.80 & 21.88 & 30.38  & 25.84  & 34.85  & 34.97  & 32.28  \\ 
          \ce{O    + CH4   ->   OH    +  CH3  }      & $V_f$ & 14.40 & 1.56  & 7.60   & 3.36   & 13.76  & 14.68  & 13.85  \\ 
                                                     & $V_b$ & 8.90  & 3.14  & 8.81   & 4.75   & 7.69   & 9.00   & 8.96   \\ 
          \ce{H    + PH3   ->   H2    +  PH2  }      & $V_f$ & 2.90  & -3.09 & -1.05  & -2.71  & 3.73   & 2.73   & 1.67   \\ 
                                                     & $V_b$ & 24.70 & 19.17 & 22.41  & 19.83  & 23.57  & 22.86  & 22.72  \\ 
          \ce{H    + OH    ->   H2    +  O    }      & $V_f$ & 10.90 & 3.02  & 8.07   & 4.61   & 11.08  & 10.90  & 10.61  \\ 
                                                     & $V_b$ & 13.20 & 1.85  & 7.83   & 3.66   & 12.85  & 13.34  & 12.70  \\ 
          \ce{H    + H2S   ->   H2    +  HS   }      & $V_f$ & 3.90  & -2.62 & -0.60  & -1.94  & 4.75   & 3.85   & 2.92   \\ 
                                                     & $V_b$ & 17.20 & 11.04 & 13.90  & 11.96  & 16.28  & 15.54  & 15.79  \\ 
          \ce{O    + HCl   ->   OH    +  Cl   }      & $V_f$ & 10.40 & -3.60 & 22.20  & 1.14   & 12.43  & 12.19  & 10.95  \\ 
                                                     & $V_b$ & 9.90  & -0.96 & 14.32  & 3.53   & 11.06  & 9.90   & 10.12  \\ 
          \ce{NH2  + CH3   ->   NH    +  CH4  }      & $V_f$ & 8.90  & 4.58  & 8.97   & 5.67   & 8.07   & 9.32   & 9.43   \\ 
                                                     & $V_b$ & 22.00 & 11.80 & 16.70  & 12.99  & 20.03  & 21.40  & 21.35  \\ 
          \ce{NH2  + C2H5  ->   NH    +  C2H6 }      & $V_f$ & 9.80  & 6.13  & 10.50  & 7.56   & 8.95   & 10.21  & 10.56  \\ 
                                                     & $V_b$ & 19.40 & 8.88  & 14.34  & 10.62  & 17.51  & 18.78  & 18.88  \\ 
          \ce{C2H6 + NH2   ->   C2H5  +  NH3  }      & $V_f$ & 11.30 & 4.48  & 8.70   & 5.96   & 10.61  & 11.00  & 11.86  \\ 
                                                     & $V_b$ & 17.80 & 12.26 & 16.12  & 13.41  & 16.80  & 17.74  & 17.85  \\ 
          \ce{NH2  + CH4   ->   NH3   +  CH3  }      & $V_f$ & 13.90 & 7.33  & 11.05  & 8.63   & 13.17  & 13.70  & 14.51  \\ 
                                                     & $V_b$ & 16.90 & 10.63 & 14.59  & 11.81  & 15.95  & 16.94  & 16.90  \\ 
          \ce{C5H8         ->   C5H8          }      & $V_f$ & 39.70 & 33.77 & 32.60  & 37.07  & 41.58  & 38.90  & 45.77  \\ 
                                                     & $V_b$ & 39.70 & 33.77 & 32.60  & 37.07  & 41.58  & 38.90  & 45.77  \\ \hline\hline
        
        \end{longtable*}
        
        \begin{longtable*}[ht]{c|rrrrrrrr}
      \caption{
      Dissociation energy (in kcal/mol) 
      with the W2-F12 energies as references (Ref.) and the calculated results of the SCAN, SCAN@HF, SCAN(SIC), xDH@PBE0, xDH@SCAN and xDH@SCAN(SIC) methods
      for the SIE4x4 database of four positively charged dimers (\ce{H2+},\ce{He2+},\ce{(NH3)2+}, and \ce{(H2O)2+}) at 5 data points are sampled at $1.5$ \AA, $2.3$ \AA, $3.1$ \AA, $4.7$ \AA, and $6.0$ \AA, while for NaCl, the data points are sampled at $2.6$ \AA, $3.8$ \AA, $4.0$ \AA, $5.6$ \AA, and $7.8$ \AA. 
      }
      \label{tb:SIE} \\
      \hline\hline
        Systems        & Lengths & Ref. & SCAN  & \SCANHF & \SCANSIC & \xDHPBE  & \xDHSCAN & \xDHSCANSIC    \\ \hline
        \ce{H2+}       & 1.00     & 64.4 & 67.18 & 67.16   & 67.15    & 64.96    & 65.34   & 65.28 \\ 
                       & 1.25     & 58.9 & 64.19 & 64.05   & 64.13    & 59.86    & 59.55   & 59.81 \\ 
                       & 1.50     & 48.7 & 56.66 & 56.50   & 56.60    & 49.83    & 48.84   & 49.69 \\ 
                       & 1.75     & 38.3 & 49.49 & 48.93   & 49.44    & 39.74    & 38.38   & 39.45 \\ \hline
        \ce{He2+}      & 1.00     & 56.9 & 75.46 & 72.59   & 74.71    & 57.42    & 57.93   & 59.40 \\ 
                       & 1.25     & 46.9 & 72.39 & 68.48   & 71.63    & 47.21    & 46.70   & 48.79 \\ 
                       & 1.50     & 31.3 & 64.42 & 59.60   & 63.67    & 32.10    & 30.90   & 34.06 \\ 
                       & 1.75     & 19.1 & 59.48 & 54.06   & 58.73    & 20.76    & 19.10   & 23.25 \\ \hline
        \ce{(NH3)2+}   & 1.00     & 35.9 & 42.72 & 40.14   & 42.69    & 36.86    & 36.34   & 36.78 \\ 
                       & 1.25     & 25.9 & 37.73 & 33.30   & 37.69    & 26.40    & 24.98   & 26.44 \\ 
                       & 1.50     & 13.4 & 30.38 & 24.38   & 30.33    & 14.03    & 12.17   & 14.38 \\ 
                       & 1.75     & 4.9  & 26.63 & 19.69   & 26.59    & 6.09     & 3.95    & 6.73 \\ \hline
        \ce{(H2O)2+}   & 1.00     & 39.7 & 52.61 & 48.41   & 52.63    & 41.38    & 40.69   & 42.00 \\ 
                       & 1.25     & 29.1 & 48.50 & 42.40   & 48.49    & 30.41    & 29.14   & 31.41 \\ 
                       & 1.50     & 16.9 & 42.18 & 34.91   & 42.15    & 18.60    & 16.90   & 19.97 \\ 
                       & 1.75     & 9.3  & 39.42 & 31.88   & 39.38    & 11.56    & 9.77    & 13.31 \\ \hline\hline
    \end{longtable*}

        \begin{longtable*}[ht]{c|rrrrrrrr}
      \caption{
        Decomposition reaction energy (in kcal/mol)
        with the CCSD(T) results as references (Ref.) and the calculated results of the SCAN, SCAN@HF, SCAN(SIC), xDH@PBE0, xDH@SCAN and xDH@SCAN(SIC) methods
        for the ALK8 database of the alkaline metal complexes into their dimers. 
      }
      \label{tb:ALK8} \\
      \hline\hline
      Reactions                       & Ref.   & SCAN   & \SCANHF & \SCANSIC & \xDHPBE & \xDHSCAN & \xDHSCANSIC \\ \hline
      \ce{Li8 -> 4Li2}	              & 86.47  & 94.83  & 91.63   & 93.55    & 87.54    & 96.23    & 81.01         \\
      \ce{Na8 -> 4Na2}                & 53.15  & 60.96  & 57.51   & 60.00    & 54.25    & 72.61    & 53.26         \\
      \ce{Li4(CH3)4 -> 4Li(CH3)}      & 131.13 & 133.43 & 133.24  & 133.26   & 128.27   & 135.84   & 141.67        \\
      \ce{Li3(CH3) -> Li(CH3) + Li2}  & 34.51  & 35.08  & 35.22   & 35.81    & 33.67    & 35.42    & 32.68         \\
      \ce{Li2(CH4) -> Li(CH3) + LiH}  & 47.42  & 47.63  & 47.87   & 47.77    & 46.83    & 48.83    & 47.91         \\
      \ce{Li5(CH) -> Li4C + LiH}      & 66.28  & 68.55  & 67.50   & 70.00    & 59.48    & 62.45    & 66.44         \\
      \ce{Li2(CH2)N2 -> 2LiCH2N}      & 56.55  & 55.17  & 55.29   & 54.72    & 55.51    & 56.52    & 60.81         \\
      \ce{Na + LiNaH2 -> Li + Na2J2}  & 25.30  & 23.14  & 24.10   & 23.25    & 27.32    & 25.13    & 26.30         \\ \hline\hline
    \end{longtable*}

    \begin{longtable*}[ht]{c|rrrrrrrr}
      \caption{
      Dissociation energy (in kcal/mol) 
      with the CCSD(T) results as references (Ref.) and the calculated results of the SCAN, SCAN@HF, SCAN(SIC), xDH@PBE0, xDH@SCAN and xDH@SCAN(SIC) methods
      for the HD2x5 database of the LiF and NaCl molecules at their 5 different bond lengthes.
      }
      \label{tb:HD2x5} \\
        \hline \hline
        Systems & Lengths(\AA) & CCSD(T) & SCAN    & \SCANHF & \SCANSIC & \xDHPBE & \xDHSCAN & \xDHSCANSIC \\ \hline
        LiF     & 1.5        & -133.68 & -133.83 & -132.11  & -130.17   & -136.74  & -141.40  & -128.07       \\
                & 2.3        & -103.86 & -99.97  & -97.51   & -97.28    & -104.23  & -103.76  & -98.85        \\
                & 3.1        & -66.49  & -61.41  & -59.99   & -64.40    & -67.31   & -61.78   & -65.82        \\
                & 4.7        & -27.48  & -27.33  & 0.02     & -27.26    & -47.66   & 19.19    & -24.71        \\
                & 6          & -10.59  & -19.18  & 0.02     & -9.09     & -93.92   & 39.57    & -10.81        \\ \hline
        NaCl    & 2.6        & -94.89  & -92.20  & -94.09   & -91.93    & -96.68   & -97.54   & -92.47        \\
                & 3.8        & -57.45  & -52.87  & -56.65   & -56.60    & -58.65   & -53.32   & -59.73        \\
                & 4          & -52.48  & -47.20  & -51.47   & -51.57    & -53.52   & -42.53   & -54.81        \\
                & 5.6        & -25.25  & -25.05  & -1.01    & -22.40    & -40.55   & 3.37     & -24.67        \\
                & 7.8        & -6.81   & -17.05  & -0.04    & -1.87     & -72.21   & 23.94    & -7.33         \\ \hline\hline
    \end{longtable*}
\end{appendix}

\end{document}